\documentclass{aa}  

\usepackage{natbib}
\usepackage{graphicx}
\usepackage{txfonts}

\usepackage{xcolor}
\usepackage{tikz}
\usetikzlibrary{bayesnet}

\usepackage{diagbox}

\newcommand{\de}{\text{d}}
\newcommand{\Msun}{\text{M}_{\odot}}
\newcommand{\kmsec}{\text{km}\,\text{s}^{-1}}
\newcommand{\kmsecsq}{\text{km}^2\,\text{s}^{-2}}
\newcommand{\kpc}{\text{kpc}}
\newcommand{\pc}{\text{pc}} 
\newcommand{\yr}{\text{yr}}
\newcommand{\Myr}{\text{Myr}}
\newcommand{\Msunppcc}{\Msun\pc^{-3}}
\newcommand{\Msunppcsquare}{\Msun\pc^{-2}}

\newcommand{\popp}{\boldsymbol{\Psi}}

\newcommand{\data}{\boldsymbol{d}}

\newcommand{\sigm}{\text{sigm}}

\usepackage{hyperref}
\begin{document}

   \title{Weighing the Galactic disk using phase-space spirals \\ I. Tests on one-dimensional simulations}

   \author{A. Widmark
          \inst{1}
          \and
          C.F.P. Laporte
          \inst{2}
          \and
          P.F. de Salas
          \inst{3}
          }

   \institute{Dark Cosmology Centre, Niels Bohr Institute, University of Copenhagen, Jagtvej 128, 2200 Copenhagen N, Denmark\\
   \email{axel.widmark@nbi.ku.dk}
   \and
   Kavli Institute for the Physics and Mathematics of the Universe (WPI), The University of Tokyo Institutes for Advanced Study (UTIAS), The University of Tokyo, Chiba 277-8583, Japan
   \and
   The Oskar Klein Centre for Cosmoparticle Physics, Department of Physics, Stockholm University, AlbaNova, 10691 Stockholm, Sweden
    }

   \date{Received Month XX, XXXX; accepted Month XX, XXXX}

 
  \abstract{We present a new method for inferring the gravitational potential of the Galactic disk, using the time-varying structure of a phase-space spiral in the $(z,w)$-plane (where $z$ and $w$ represent vertical position and vertical velocity). Our method of inference extracts information from the shape of the spiral and disregards the bulk density distribution that is usually used to perform dynamical mass measurements. In this manner, it is complementary to traditional methods that are based on the assumption of a steady state. Our method consists of fitting an analytical model for the phase-space spiral to data, where the spiral is seen as a perturbation of the stellar number density in the $(z,w)$-plane. We tested our method on one-dimensional simulations, which were initiated in a steady state and then perturbed by an external force similar to that of a passing satellite. We were able to retrieve the true gravitational potentials of the simulations with high accuracy. The gravitational potential at 400--500 parsec distances from the disk mid-plane was inferred with an error of only a few percent. This is the first paper of a series in which we plan to test and refine our method on more complex simulations, as well as apply our method to \emph{Gaia} data.}

   \keywords{Galaxy: kinematics and dynamics -- Galaxy: disk -- solar neighborhood -- Astrometry}

   \maketitle
%

\section{Introduction}\label{sec:intro}

Determining the gravitational potential of the Milky Way is important for constraining its composition, history, and dynamical properties \citep{1998MNRAS.294..429D, Klypin:2001xu, Widrow:2008yg,2010A&A...509A..25W,McMillan:2011wd,2014ApJ...794...59K,2017MNRAS.465..798C,2017MNRAS.465...76M,2020MNRAS.494.6001N,2020MNRAS.494.4291C,2020ApJ...894...10L}. Weighing the Galaxy is especially important for understanding dark matter and its spatial distribution. The local density of dark matter influences the sensitivity of direct and indirect detection experiments \citep{Jungman:1995df,2015PrPNP..85....1K} and could potentially depend on dark substructures, which would inform us of its particle nature \citep{10.1111/j.1365-2966.2008.13643.x,0004-637X-703-2-2275,Fan:2013tia,2014MNRAS.444..515R}.

The first dynamical mass measurements of the Galactic disk were carried out roughly a century ago, by \cite{Kapteyn1922}, \cite{10.1093/mnras/82.3.122}, and \cite{Oort1932}. They were able to obtain remarkably accurate estimates of the total matter density in the solar neighbourhood, under the assumption that the local population of stars was in a stable configuration. Modern methods use the same basic principles: By assuming a steady state, the stellar number density distribution and velocity distribution are interrelated via the gravitational potential (although there are also recent developments in terms of probing the gravitational potential with direct acceleration measurements, for example using pulsar timings; \citealt{chakrabarti21}). With the advent of the \emph{Gaia} mission \citep{2016A&A...595A...1G}, which recently published an early instalment of its third data release (EDR3; \citealt{2020arXiv201201533G}), the stars of our Galaxy are observed with greater depth and precision than ever before.

One great discovery directly enabled by the \emph{Gaia} survey is that of the now famous phase-space spiral \citep{2018Natur.561..360A}. This spiral is visible in the plane of vertical position and vertical velocity (where vertical refers to the direction perpendicular to the Galactic disk), either as a function of the median velocity in the azimuthal direction or as a perturbation to the stellar number density. In terms of the stellar number density in the plane of vertical position and velocity, the phase-space spiral can be seen by eye only at higher vertical energies; however, \cite{2019MNRAS.485.3134L} demonstrated that the spiral, when plotted as a relative over- and under-density with respect to a bulk density component, is also visible at lower vertical energies.

The phase-space spiral is a clear example that the Galaxy is actually not in a steady state; other examples are ridges in stellar number counts as a function of azimuthal velocity and galactocentric radius \citep{2018Natur.561..360A} and pre-\emph{Gaia} studies of the solar neighbourhood \citep{2009MNRAS.396L..56M,2012ApJ...750L..41W} and the Milky Way disk-halo interface \citep{2002ApJ...569..245N,2003astro.ph..7258R,2015ApJ...801..105X, 2015MNRAS.452..676P, 2018Natur.555..334B}. It remains an open question as to what extent these structures can bias dynamical mass measurements (see for example \citealt{2019ApJ...879L..15H} and \citealt{2020A&A...643A..75S}). However, because \emph{Gaia} is precise enough to resolve and identify such features, it seems reasonable that we have already entered an era where time-varying dynamical structures induce significant systematic biases that in many cases surpass the strictly statistical uncertainty of traditional methods. While most local mass measurements quote rather small statistical uncertainties, there are significant discrepancies between studies, depending for example on the choice of method, and sometimes even within studies, depending for example on the choice of stellar tracer population (see reviews by \citealt{Read2014} and \citealt{2020arXiv201211477D} on recent determinations of the local dark matter density). Furthermore, in a study by \cite{2020arXiv201102490W}, in which the total matter density is measured in subregions of the solar neighbourhood, the authors argue that their inferred matter density distributions can only be explained by biases induced by time-varying dynamics.

This work is an attempt to move beyond the traditional methods that are based on the assumption of a steady state. We demonstrate that the gravitational potential can be inferred from the shape of a phase-space spiral in the plane of vertical position and vertical velocity. Our method, as formulated in this paper, uses the relative stellar number density perturbation of the spiral with respect to a smooth background component. This background component, which we refer to as the bulk, is the quantity that is traditionally used to perform dynamical mass measurement. Our method of inference extracts information only from the shape of the spiral, while the bulk density distribution is disregarded and does not influence, nor is influenced by, the inferred gravitational potential. We tested our method on one-dimensional simulations, which were initiated in a steady state and then perturbed by applying an external force similar to that of a passing satellite. The gravitational potential was inferred with high accuracy, illustrating that time-varying structures are not necessarily obstacles to be overcome, but can be regarded as valuable assets containing useful information.

This article is structured in the following way. In Sect.~\ref{sec:model} we describe our analytical model for the phase-space spiral, as well as our statistical model of inference. In Sect.~\ref{sec:simulations} we present our simulation on which we test our model of inference. The results of those tests are found in Sect~\ref{sec:results}. In Sects.~\ref{sec:discussion} and \ref{sec:conclusion} we discuss our results and conclude.

\section{Analytical model of the spiral}\label{sec:model}

In this work we considered a simple analytical model for the Galactic disk phase-space spiral. This analytical model was based on the three following simplifying assumptions: (i) We reduced the problem to a single spatial dimension and considered motion only in the direction perpendicular to the Galactic plane.
(ii) The phase-space spiral was modelled as a first order perturbation of the stellar phase-space density distribution, inhabiting a stationary gravitational potential. As such, our analytical model neglects the self-gravity of the spiral structure. (iii) The initial perturbation is not in the shape of a spiral to begin with; it is modelled as a function that depends on, and is separable with respect to, the total vertical energy and angle of vertical oscillation (see Sect.~\ref{sec:definitions} for the mathematical definition of these quantities).

The actual phase-space spiral of the Milky Way is of course not subject to these assumptions. Rather, it inhabits the Galaxy's three-dimensional potential and six-dimensional phase-space structure, it is subject to self-gravity, and the initial perturbation could have a more complex structure (the Galactic disk might be subject to several or even continuous spiral-producing perturbations; see for example \citealt{2018MNRAS.481.3794H,2018MNRAS.481.1501B,2019MNRAS.485.3134L,2019A&A...622L...6K}). A discussion on the validity of these assumption is provided in  Sect.~\ref{sec:discussion}.

\subsection{Coordinate system and stellar kinematics}\label{sec:definitions}
In order to describe the analytical model for a phase-space spiral, we begin by making the following definitions.

The vertical position, written as $z$ and also referred to as height, is defined in relation to the disk's mid-plane, such that the gravitational potential, written as $\Phi(z)$, has a zero-valued minimum at $z=0~\pc$. The vertical velocity, written as $w$, is defined in relation to the disk's rest frame. Hence, a star's vertical energy\footnote{This quantity is, strictly speaking, in units of vertical energy per mass but is referred to as vertical energy throughout this article.} is equal to
\begin{equation}\label{eq:E_z}
    E_z = \Phi(z) + \frac{w^2}{2}.
\end{equation}
Given a fixed value for $E_z$, the absolute value of $w$ is fully determined by the absolute value of $z$, and vice versa.

In reality, stars are not observed from the rest frame of the disk itself, but rather from the vantage point of the Solar system. The disk frame coordinates ($z$, $w$) and solar frame coordinates ($Z$, $W$) have the relation
\begin{equation}\label{eq:sun_coords}
\begin{split}
    z & = Z + Z_\odot, \\
    w & = W + W_\odot,
\end{split}
\end{equation}
where $Z_\odot$ is the height of the Sun with respect to the disk's mid-plane and $W_\odot$ is the vertical velocity of the Sun in the disk's rest frame.

A star with a certain vertical energy will reach a maximum height, $z_\text{max}$, which fulfils that
\begin{equation}\label{eq:z_max}
    E_z = \Phi(z_\text{max}).
\end{equation}
The star's vertical oscillation has a total time period of
\begin{equation}\label{eq:period}
    P(E_z) = \oint \frac{\de z}{w} =  4\int_0^{z_\text{max}} \frac{\de z}{\sqrt{2[E_z-\Phi(z)]}},
\end{equation}
where the factor 4 on the right-hand side is due to the integral only covering one of four quadrants of the oscillation in the $(z,w)$-plane.

We define the starting point of the vertical oscillation as passing through the disk's mid-plane with a positive vertical velocity ($z=0$, $w>0$). Finally, we define an angle of oscillation ($\varphi$) as the fraction of elapsed time relative to the total period, where a complete period amounts to $2\pi$, such that $[z,w](\varphi) = [z,w](\varphi+2\pi n)$, where $n$ is an integer. The angle, here chosen to lie in the range $[0,2\pi)$, can be calculated according to
\begin{equation}\label{eq:angle_of_z}
\varphi(z,w) = 
\begin{cases}
    2 \pi P^{-1}{\displaystyle\int_0^{|z|}} \dfrac{\de z'}{\sqrt{2[E_z-\Phi(z')]}} & \text{if}\,z\geq0\,\text{and}\,w\geq0, \\
    \pi - 2 \pi P^{-1}{\displaystyle\int_0^{|z|}} \dfrac{\de z'}{\sqrt{2[E_z-\Phi(z')]}} & \text{if}\,z\geq0\,\text{and}\,w<0, \\
    \pi + 2 \pi P^{-1}{\displaystyle\int_0^{|z|}} \dfrac{\de z'}{\sqrt{2[E_z-\Phi(z')]}} & \text{if}\,z<0\,\text{and}\,w<0, \\
    2\pi - 2 \pi P^{-1}{\displaystyle\int_0^{|z|}} \dfrac{\de z'}{\sqrt{2[E_z-\Phi(z')]}} & \text{if}\,z<0\,\text{and}\,w\geq0,
\end{cases}
\end{equation}
where these four cases correspond to the four quadrants of the $(z,w)$-plane and it is implicit that $E_z$ is given by Eq~\eqref{eq:E_z} and $P$ by Eq.~\eqref{eq:period}.\footnote{In Eq.~\eqref{eq:angle_of_z}, the integral is written with its upper bound in absolute value ($|z|$) in order to make the expression more explicit in terms of its sign (making the integral itself a positive quantity). This formulation is contingent on the potential being mirror-symmetric with respect to the Galactic plane.}

In our analytical model, we parametrise the gravitational potential according to
\begin{equation}\label{eq:phi_parametrisation}
    \Phi(z  \, | \, \rho_h) = \sum_{h=1}^{4}
    \frac{4 \pi G \rho_h}{(2^{h-1} \times 100~\pc)^2}\log\Bigg[\cosh\Bigg(\dfrac{z}{2^{h-1} \times 100~\pc}\Bigg)\Bigg].
\end{equation}
Via the one-dimensional Poisson equation,
\begin{equation}\label{eq:Poisson}
        \frac{\partial^2\Phi(z)}{\partial z^2} = 4\pi G \rho(z),
\end{equation}
this corresponds to a total matter density distribution equal to
\begin{equation}\label{eq:rho_parametrisation}
    \rho(z) = \sum_{h=1}^{4} \rho_h \cosh^{-2}\Bigg(\dfrac{z}{2^{h-1} \times 100~\pc}\Bigg),
\end{equation}
where the four separate components labelled by the index $h$ have scale heights of $\{100,200,400,800\}~\pc$, respectively. This functional form allows us to emulate, with flexibility, the general characteristics of the total matter density of the Galactic disk in the solar neighbourhood (in terms of total mid-plane density, scale height, and shape; see for example \citealt{2015ApJ...814...13M} and \citealt{Schutz:2017tfp}).

\subsection{Spiral winding}\label{sec:winding}

In our analytical model, the initial perturbation was assumed to be an over-density along some initial angle ($\tilde{\varphi}_0$). For example, the over-density could lie along a vertical or horizontal semi-axis in the $(z,w)$-plane.

The angle of a star vertically oscillating in the disk evolves with time according to
\begin{equation}\label{eq:angle_of_time}
    \tilde{\varphi}(t,E_z) = \tilde{\varphi}_0 + 2\pi\frac{t}{P(E_z)},
\end{equation}
where the period $P(E_z)$ has an implicit dependence on $\Phi$.

Because the initial over-density is assumed to be described by a single initial angle, which we write as $\tilde{\varphi}_0$, this equation also describes the winding of the phase-space spiral and how it evolves with time. The winding of the spiral arises from variations in the period with respect to $E_z$. If the gravitational potential of the disk were harmonic, such that the period were invariant with respect to $E_z$, no winding would occur. However, the matter density of the Galactic disk decreases with distance from the mid-plane, such that the period of oscillation increases with $E_z$. As a result, an initial perturbation will wind into a spiral that is dragging if considered from the inside out -- in other words, winding in the opposite direction of the rotation of the stars (which evolve clockwise)  in the $(z,w)$-plane.

\subsection{Spirals represented as line graphs}\label{sec:examples}

The shape of the phase-space spiral in the $(z,w)$-plane is determined by: the initial angle, $\tilde{\varphi}_0$; the time since the initial perturbation, $t$; and the gravitational potential, $\Phi(z)$. Here, we consider a simplistic case where the spiral is a one-dimensional curve in the $(z,w)$-plane. This curve can be written in terms of its angle as a function of vertical energy, $\tilde{\varphi}(E_z)$, which is a smooth and strictly decreasing function. The angle can in turn be translated into phase-space coordinates, $z[\tilde{\varphi}(E_z)]$ and $w[\tilde{\varphi}(E_z)]$, according to the equations presented in Sects.~\ref{sec:definitions} and \ref{sec:winding}.

\begin{figure}
        \includegraphics[width=1.\columnwidth]{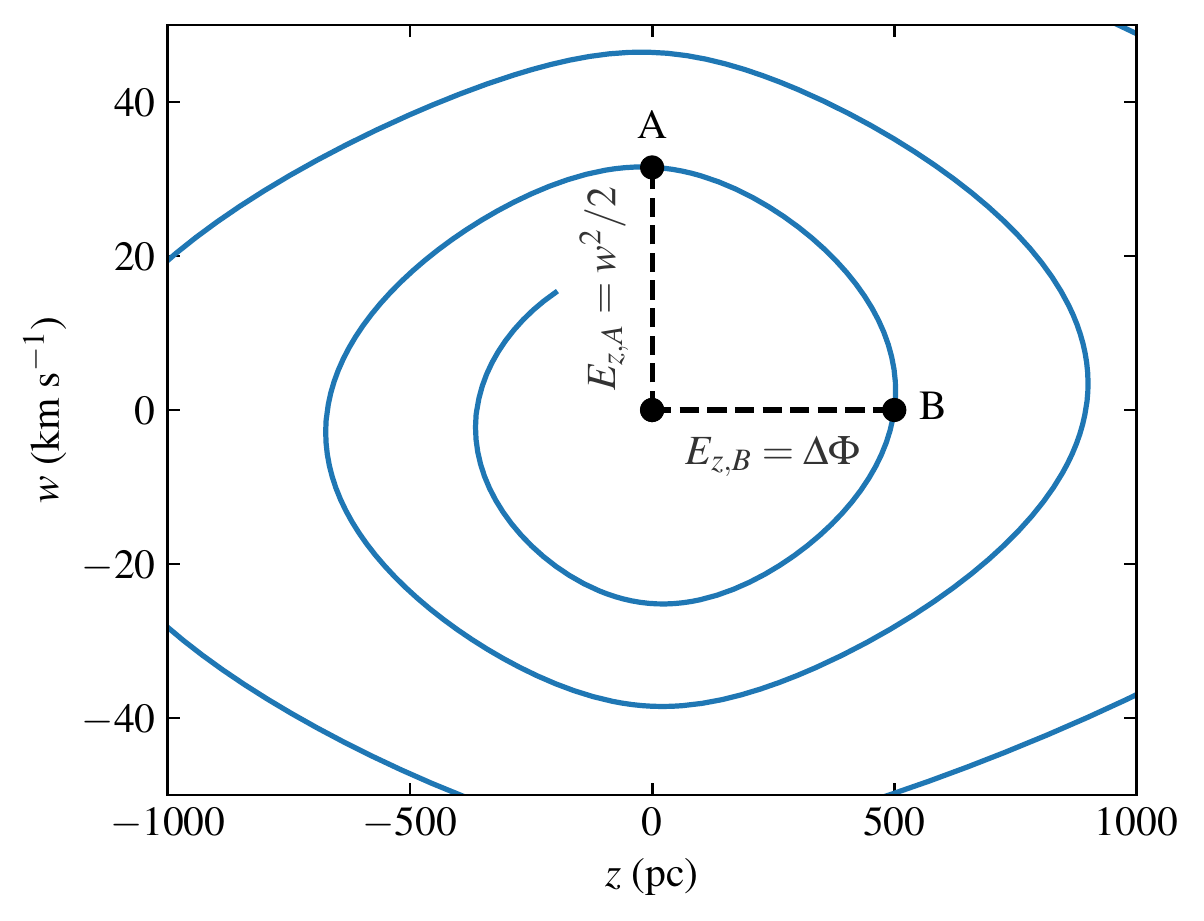}
    \caption{Phase-space spiral represented as a line graph, shown in solid blue. Two points, labelled A and B, are connected to the origin point via dashed black lines. Points A and B are both associated with a total vertical energy: $E_{z,A} = w^2/2$ is given by the vertical velocity where the spiral passes through the Galactic mid-plane; and $E_{z,B} = \Phi(z)$ is given by gravitational potential at a point of the spiral with zero velocity. In the limit of high winding, the difference $E_{z,A}-E_{z,B}$ would approach zero from above.}
    \label{fig:schematic}
\end{figure}

An example of such a spiral is presented in Fig.~\ref{fig:schematic}. For this spiral, the time since the perturbation is $t=600~\Myr$, the initial angle is chosen such that the spiral cuts the $z$-axis at $500~\pc$, and the gravitational potential is determined via Eq.~\eqref{eq:phi_parametrisation} with parameters $\rho_h = \{0,\,0.06,\,0.03,\,0\}~\Msunppcc$.

Any point on the spiral in Fig.~\ref{fig:schematic} is associated with a specific vertical energy. We have highlighted points A and B, for which the vertical energy takes the form of either purely kinematic or potential energy:
\begin{equation}
\begin{split}
    E_{z,A} & = \frac{w^2}{2} =  \frac{(31.5 ~\kmsec)^2}{2} = 496 ~\kmsecsq, \\
    E_{z,B} & = \Delta \Phi = \Phi(500~\pc) = 400 ~\kmsecsq.
\end{split}
\end{equation}
In the limit of high winding, the difference in energy between these two points will be small, allowing us to make measurements of the disk's mass knowing only the $(z,w)$ coordinates of the two points. In the specific example above, the numerical values for the two energies do have a sizeable difference. However, even in a case like this, where the spiral's winding is not very high, the gravitational potential can be robustly and accurately inferred by considering the spiral curve over a longer segment, for example over half a period. The shape of the spiral will be strongly constrained by the gravitational potential and from the fact that $\tilde{\varphi}(E_z)$ is a smooth function.

\begin{figure}
        \includegraphics[width=1.\columnwidth]{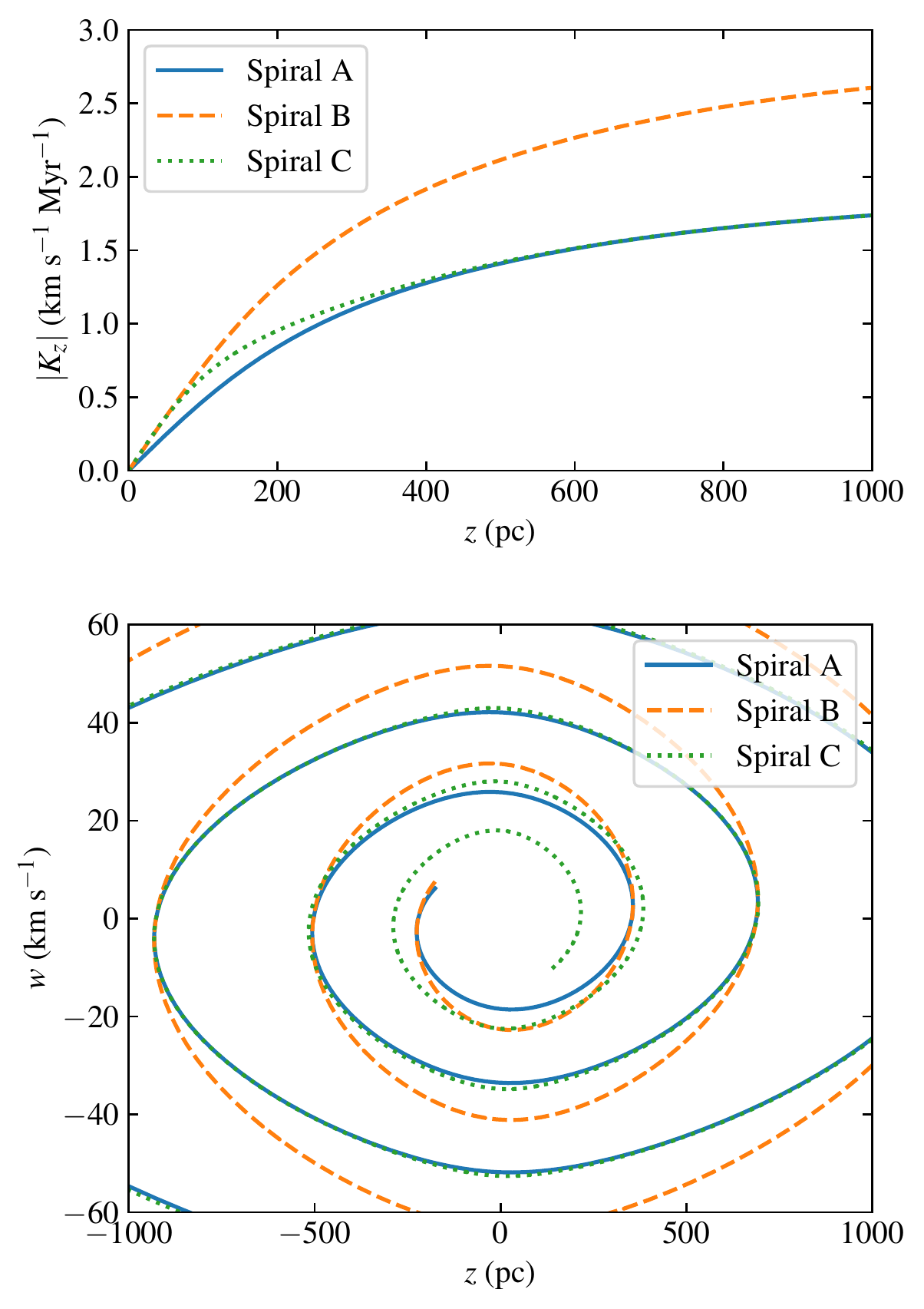}
    \caption{Three phase-space spirals (labelled A, B, and C) inhabiting three different gravitational potentials. The top panel shows the vertical force per mass ($K_z = - \partial \Phi / \partial z$) as a function of height for the respective gravitational potentials. The bottom panel shows the three spirals in the ($z,w$)-plane, where their winding time ($t$) and initial angle ($\tilde{\varphi}_0$) are chosen such that the spirals have the same phase ($\tilde{\varphi}$) and phase derivative ($\partial \tilde{\varphi} / \partial z_\text{max}$) when their vertical energy is equal to $E_z = \Phi(800~\pc)$.}
    \label{fig:three_spirals}
\end{figure}

In order to illustrate how the shape of the spiral can change when varying the gravitational potential, we show three different examples in Fig.~\ref{fig:three_spirals}. In the upper panel of this figure, the gravitational potentials are illustrated in terms of their vertical force per mass ($K_z = - \partial \Phi / \partial z$). The gravitational potential of spiral A corresponds to a matter density distribution with parameters $\rho_{h,A} = \{0,\,0.05,\,0.02,\,0.02\}~\Msunppcc$.
The matter density distribution of spiral B is identical to that of spiral A in terms of shape, but multiplied by a factor of 1.5, giving $\rho_{h,B} = \{0,\,0.075,\,0.03,\,0.03\}~\Msunppcc$. Finally, the matter density distribution of spiral C has parameters $\rho_{h,C} = \{0.1,\,0,\,0.02,\,0.02\}~\Msunppcc$. This is identical to that of spiral A for $z\gtrsim 400~\pc$ (also in terms of the gravitation force) but differs at lower heights.

The three spirals, shown in the bottom panel of Fig.~\ref{fig:three_spirals}, are individually normalised in terms of $t$ and $\tilde{\varphi}_0$ to have the same winding behaviour for $E_z = \Phi(800~\pc)$. This makes, in some sense, the three spirals maximally similar to one another. Despite this, the three spirals can be differentiated from one another as a result of inhabiting different gravitational potentials. In terms of shape, spirals A and B are identical; they differ solely by a factor of $\sqrt{1.5}$ in their velocity coordinate, $w$. Spirals A and C are practically identical when $z \gtrsim 400~\pc$ and $E_z \gtrsim \Phi(400~\pc)$. They differ slightly when $z \simeq 0$ and $E_z \gtrsim \Phi(400~\pc)$ due to spiral C inhabiting a potential that is slightly steeper close to the mid-plane. In the innermost region of the plot, where $E_z \lesssim \Phi(400~\pc)$, spiral C also differs from spiral A in terms of its winding behaviour because the oscillation period differs significantly for low $z_\text{max}$ and $E_z$.

The examples presented above illustrate the general principles of how the gravitational potential affects the phase-space spiral shape, and in effect how the latter can be used to infer the former. The total amount of winding is affected by the total mass of the disk as well as $t$, so in this sense these quantities are degenerate. However, as we see in the comparison between spirals A and B, for which the total mass differs by a factor $1.5$ and $t$ differs by $1/\sqrt{1.5}$, the shape of the spirals are dramatically different despite having the same amount of winding in terms of $\partial \tilde{\varphi} / \partial z_\text{max}$. By comparing spirals A and C, we also see that the spiral shape in a certain segment, which we can define by some range in $E_z$, is sensitive to the gravitational potential at the associated energy. For example, the gravitational potential close to the mid-plane affects the spiral's winding behaviour, especially at low vertical energies.

\subsection{Complete spiral model}

In this work we fitted an analytical phase-space spiral to data coming from a one-dimensional simulation of the Galactic disk. In our complete analytical model of inference, the phase-space spiral was not modelled as the idealised one-dimensional structure discussed in Sect.~\ref{sec:examples}, but instead as a continuous function with respect to $(z,w)$. This full model is described here.

The free parameters of our model of inference are listed in Table~\ref{tab:model_parameters}. The free parameters, encapsulated in the quantity $\popp$, are split into three subgroups that determine the bulk phase-space density ($\popp_\text{bulk}$), the relative phase-space density perturbation of the spiral ($\popp_\text{spiral}$), and the current phase-space position of the Sun ($\popp_\odot$). These model components are described in detail below.

{\renewcommand{\arraystretch}{1.6}
\begin{table*}[ht]
        \centering
        \caption{Free parameters in our analytical model of inference. These parameters are split into three groups, $\popp = \{\popp_\text{bulk}, \popp_\text{spiral}, \popp_\odot\}$, which correspond to: (i) the parameters that determine the bulk phase-space density; (ii) the relative phase-space density perturbation of the spiral; and (iii) the solar phase-space parameters. The total number of free parameters is equal to $10+3K$, where $K$ is the number of Gaussian components in the bulk stellar density.}
        \label{tab:model_parameters}
    \begin{tabular}{| l | l |}
                \hline
                $\popp_\text{bulk}$  & Bulk phase-space density parameters \\
                \hline
                $a_k$ & Weights of the bulk density Gaussian mixture model \\
                $\sigma_{z,k}$, $\sigma_{w,k}$ & Dispersions of the bulk density Gaussian mixture model \\
                \hline
                \hline
                $\popp_\text{spiral}$  & Spiral phase-space density parameters \\
                \hline
                $\rho_h$ & Matter density parameters that determine the gravitational potential \\
                $t$ & Time since the perturbation was produced \\
                $\tilde{\varphi}_0$ & Initial angle of the perturbation \\
                $\alpha$, $\beta$ & Amplitudes of the anti-symmetric and symmetric spiral components \\
                \hline
                \hline
                $\popp_\odot$  & Solar phase-space parameters \\
                \hline
                $Z_\odot$ & Height of the Sun above the Galactic plane \\
                $W_\odot$ & Vertical velocity of the Sun in the Galactic disk rest frame  \\
        \hline
        \end{tabular}
\end{table*}}

The spiral is modelled as a relative perturbation with respect to a bulk tracer stellar density. We modelled the bulk density in the $(z,w)$-plane as a sum of bivariate Gaussians labelled by the index $k$, which are all centred on $(z,w)=(0,0)$ and have diagonal covariance matrices, according to
\begin{equation}\label{eq:bulk_density}
    B(z,w\,|\,\popp_\text{bulk}) =
    \sum_k a_k \,
    \mathcal{N}\Bigg(
    \begin{bmatrix}
        z \\
        w \\
    \end{bmatrix}, \;
    \begin{bmatrix}
        \sigma_{z,k}^2 & 0  \\
        0 & \sigma_{w,k}^2 \\
    \end{bmatrix}
    \Bigg),
\end{equation}
where $\popp_\text{bulk}$ includes the Gaussian weights $a_k$ as well as the dispersions $\sigma_{z,k}$ and $\sigma_{w,k}$. The function
\begin{equation}\label{eq:multivariate_Gaussian}
        \mathcal{N}(\boldsymbol{p},\boldsymbol{\Sigma}) \equiv
    \frac{\exp\left(-\dfrac{1}{2} \boldsymbol{p}^\top\boldsymbol{\Sigma}_{\boldsymbol{p}}^{-1}\boldsymbol{p} \right)}{2\pi \sqrt{| \boldsymbol{\Sigma}_{\boldsymbol{p}} |}}
\end{equation}
is a bivariate Gaussian.

The bulk stellar density is the quantity that is usually used in order to perform dynamical mass measurements, under the assumption that this distribution is in a steady state. In the method used in this work, the bulk is merely a background that needs to be subtracted in order to extract the spiral structure. The bulk density, as expressed in Eq.~\eqref{eq:multivariate_Gaussian}, is not required to fulfil the stationary Boltzmann equation or the Jeans equations; it is completely free to vary in terms of its free parameters (meaning weights and dispersions), regardless of the gravitational potential. In this manner, the gravitational potential is completely uninformed and independent of the bulk stellar density and will be inferred only from the shape of the spiral.

The phase-space spiral was modelled as a sum of an anti-symmetric and a symmetric component, corresponding to a single- or double-armed spiral, parametrised by amplitudes $\alpha$ and $\beta$ according to
\begin{equation}\label{eq:spiral_rel_density}
\begin{split}
    S(z,w\,|\,\popp_\text{spiral}) =
    \alpha & \cos\Big[ \varphi(z,w\,|\, \rho_h)-\tilde{\varphi}(z,w,\rho_h,t, \tilde{\varphi}_0) \Big] + \\
    \beta & \cos\Big\{ 2\big[ \varphi(z,w\,|\, \rho_h)-\tilde{\varphi}(z,w,\rho_h,t, \tilde{\varphi}_0) \big]\Big\}.
\end{split}
\end{equation}
Here, $\varphi(z,w\,|\,\rho_h)$ is the intrinsic phase of the $(z,w)$ coordinates according to Eq.~\eqref{eq:angle_of_z}, although here it is written with an explicit dependence on the parameters that determine the gravitational potentia, and $\tilde{\varphi}(z,w,\Phi,t,\tilde{\varphi}_0)$ is the phase associated with the spiral according to Eq.~\eqref{eq:angle_of_time}. Because the spiral density is a relative density, the quantities $\alpha$ and $\beta$ are unit-less and constrained to fulfil that $(\alpha,\beta)>0$ and $\alpha+\beta<1$. The parameters $\rho_h$ determine the gravitational potential according to Eq.~\eqref{eq:phi_parametrisation}.

In the limit of low vertical energies, the self-gravity of the phase-space spiral is not negligible, and the spiral structure in this innermost region does not form (as found by \citealt{2019MNRAS.484.1050D}). For this reason, we imposed a smooth inner boundary to the relative spiral density, corresponding to $E_z = \Phi(400~\pc)$, within which there is no spiral. This inner boundary is defined by
\begin{equation}\label{eq:inner_boundary}
    m(z,w \, | \, \rho_h) = \text{sigm} \Bigg[
    \frac{E_z(z,w \, | \, \rho_h)-\Phi(400~\pc \, | \, \rho_h)}{\Phi(400~\pc \, | \, \rho_h)-\Phi(380~\pc \, | \, \rho_h)} \Bigg],
\end{equation}
where
\begin{equation}\label{eq:sigmoid}
    \sigm(x) \equiv \frac{1}{1+\exp(-x)}
\end{equation}
is a sigmoid function that outputs values in the range $(0,1)$.

The total phase-space density of our analytical model is equal to
\begin{equation}\label{eq:total_density}
\begin{split}
    f(z,w\,&|\,\popp_\text{bulk},\popp_\text{spiral}) = B(z,w\,|\,\popp_\text{bulk}) \\
    & \times \Big[ 1 + m(z,w \, | \, \rho_h)\, S(z,w\,|\,\popp_\text{spiral}) \Big],
\end{split}
\end{equation}
where $\popp_\text{bulk} = \{ a_k,\sigma_{z,k},\sigma_{w,k} \}$ is the subset of parameters that affect the bulk density and $\popp_\text{spiral} = \{ \rho_h,t,\tilde{\varphi}_0,\alpha,\beta \}$ are the parameters that determine the relative phase-space perturbation of the spiral.

\subsection{Data histogram and mask function}

In the model of inference, the bulk and spiral densities are fitted to data, where the data have the form of a two-dimensional histogram in the $(Z,W)$-plane of the solar frame coordinates. This histogram is the number count of observed stars in bins of size $(20~\pc)\times(1~\kmsec)$, written as $\data_{i,j}$, where $(i,j)$ labels the respective bins.

\begin{figure}
        \includegraphics[width=1.\columnwidth]{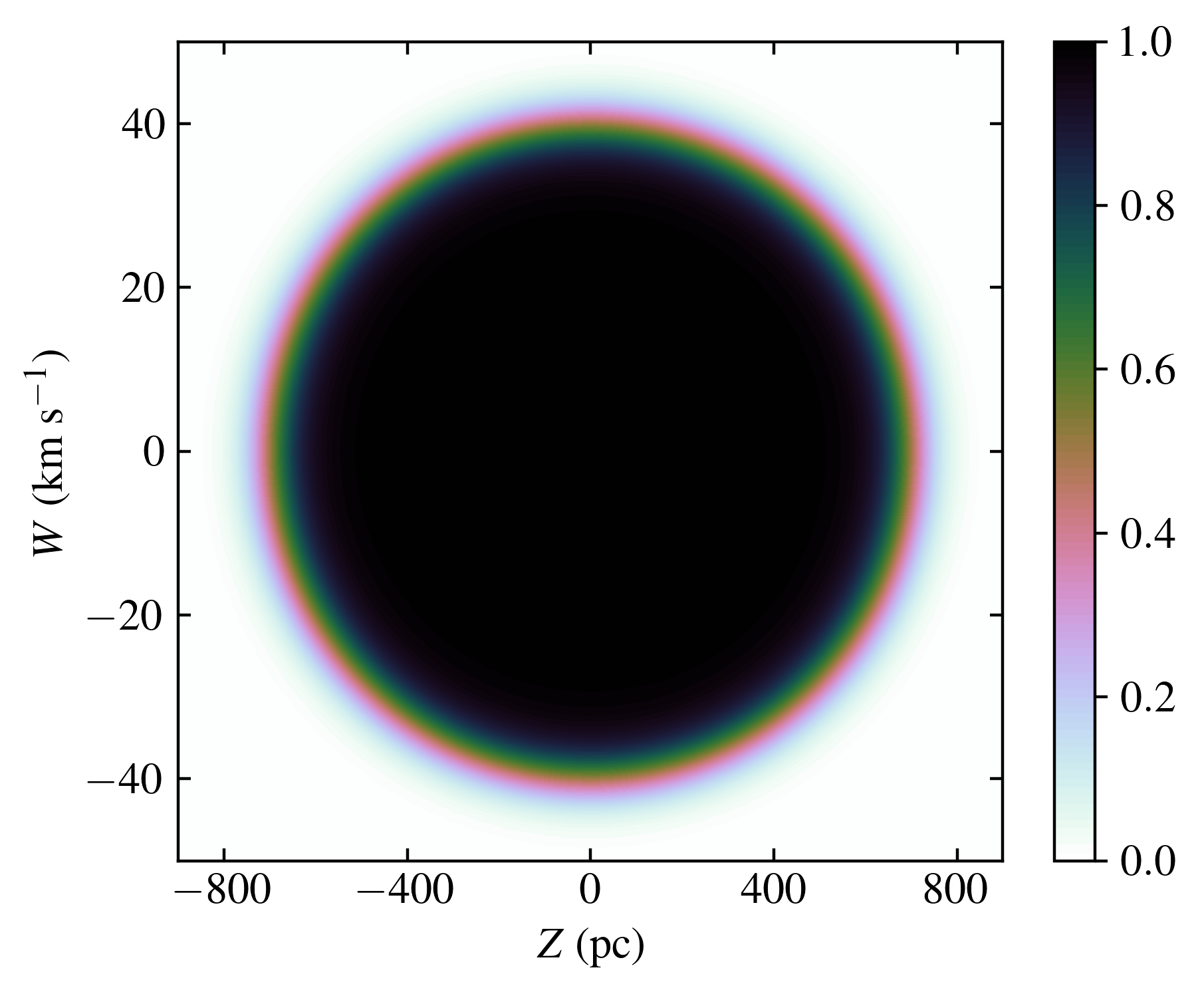}
    \caption{Mask function $M(Z,W)$ in the $(Z,W)$-plane, with boundary values $Z_\text{lim.} = 700~\pc$ and $W_\text{lim.} = 40~\kmsec$. See Eq.~\eqref{eq:mask} for details.}
    \label{fig:mask}
\end{figure}

The analytical model is fitted to a circular area of this two-dimensional histogram, defined by a mask function in the $(Z,W)$-plane with smooth outer boundaries, roughly corresponding to a constant $E_z$. The mask function is defined as
\begin{equation}\label{eq:mask}
    M(Z,W) =
    \text{sigm} \Bigg\{ 10\, \Bigg[\Bigg(\frac{Z}{Z_\text{lim.}}\Bigg)^2 + \Bigg(\frac{W}{W_\text{lim.}}\Bigg)^2 - 1 \Bigg] \Bigg\},
\end{equation}
where sigm is the sigmoid function defined in Eq.~\eqref{eq:sigmoid}. The mask function is visible in Fig.~\ref{fig:mask}.

The reason for applying such a mask has to do with the outer boundaries for which the analytical spiral model can be assumed to be a valid description. In the high energy limit, the number density of stars decreases and the spiral becomes less pronounced. Furthermore, there are additional issues that could cause systematic errors in the case of the actual Milky Way. The simplifying assumption of only considering the vertical dimension will be problematic, especially for stars with high vertical energies, which have more complex dynamical behaviour.

\subsection{Likelihood and fitting procedure}

The likelihood is given by the Poisson count comparison of the model and data of the respective bins, in the circular area defined by the mask function. The logarithm of the likelihood is equal to
\begin{equation}\label{eq:likelihood}
\begin{split}
    \ln\, & \mathcal{L}(\data_{i,j}\,|\,\popp) = \\
    & -\sum_{i,j}
    M(Z_i,W_j) \, \dfrac{[\data_{i,j}-f(Z_i+Z_\odot,W_j+W_\odot,\popp)]^2}{2 f(Z_i+Z_\odot,W_j+W_\odot,\popp)} \\
    & + \{\text{constant term}\}.
\end{split}
\end{equation}
We sought to minimise this likelihood in our fit, which would allow us to neglect any constant term in the above expression.

Our model was fitted in two separate steps. In the first step we performed a joint fit of the bulk density and solar parameters to the data while disregarding the spiral; equivalently stated, we minimised Eq.~\eqref{eq:likelihood} with respect to $\popp_\text{bulk}$ and $\popp_\odot$, while $(\alpha,\beta)=0$ remained fixed. In a second step, the bulk density and solar parameters remained fixed, while we fitted the relative density of the phase-space spiral; in other words, we minimised Eq.~\eqref{eq:likelihood} with respect to $\popp_\text{spiral}$, while $\popp_\text{bulk}$ and $\popp_\odot$ remained fixed. Only in the second step of this process, where the spiral is fitted, does the gravitational potential vary and affect the likelihood.

In order to avoid any fitting artefacts with regards to the boundary of the mask function, defined in Eq.~\eqref{eq:mask}, the two steps of the fitting procedure use slightly different masks. For the first step, where the bulk density is fitted, we used a larger mask, with $Z_\text{lim.} = 800~\pc$ and $W_\text{lim.} = 44~\kmsec$. For the second step, where the relative stellar density of the spiral is fitted, we used a smaller mask, with $Z_\text{lim.} = 700~\pc$ and $W_\text{lim.} = 40~\kmsec$.

\subsection{Model implementation}\label{sec:implementation}

The first step of the minimisation procedure is straightforward and computationally fast as it only requires fitting a two-dimensional Gaussian mixture model to data. However, the second step, where the spiral is fitted, is significantly more expensive. Calculating the spiral likelihood of Eq.~\eqref{eq:likelihood} requires calculating the intrinsic angle $\varphi(Z_i,W_j)$ as a numerical integral for each bin of the two-dimensional histogram, iterating over $(i,j)$. This is done for each individual step of the minimisation algorithm.

In order to make the algorithm computationally tractable, the method was implemented in \textsc{TensorFlow}, allowing for efficient minimisation using the Adam optimiser \citep{adamopt}. Minimising the spiral likelihood function is still computationally demanding and requires several hundred CPU hours.

\section{Simulations}\label{sec:simulations}

In this section we describe the simulations on which we tested our model of inference. The simulations are one-dimensional, constrained to the $(z,w)$-plane. The phase-space distribution of the simulations were initiated in a steady state and then perturbed by an external force, creating phase-space over- and under-densities that wind into a spiral with time. We tested our method on two simulations, labelled simulation A and simulation B.

The simulations were run with $10^5$ equally massive particles, labelled by the index $i$, representing stars and gas of the Galactic disk. Because the simulations are only in one dimension, the particles can be thought of as sheets of mass extending in the directions parallel to the Galactic disk. In the simulations, the gravitational acceleration is equal to
\begin{equation}\label{eq:vertical_force}
    K_z(z) = \frac{\de w}{\de t} =  - 4 \pi G
    \Bigg[ \rho_\text{DM} \, z + 
    \sum_i m \, \text{sign}(z-z_i) \Bigg],
\end{equation}
where $\rho_\text{DM}$ is a fixed constant representing a constant dark matter density, $m$ is the surface mass per particle (in units $\Msunppcsquare$), $z_i$ is the height of the $i$th particle, and
\begin{equation}
    \text{sign}(x) =
    \begin{cases}
        1 & \text{if } x > 0, \\
        0 & \text{if } x = 0, \\
        -1 & \text{if } x < 0.
    \end{cases}
\end{equation}

\subsection{Initial state}

The simulation particles were drawn from three separate distributions, representing stars of the thin and thick disk, as well as a component of cold gas. The initial configuration of the particles was in a steady state, and the three respective components were assumed to be isothermal. The component of halo dark matter was not represented by dynamical particles in the simulation, but by the fixed constant $\rho_\text{DM}$. The three dynamical components are defined by their respective velocity dispersions ($\sigma_{w,\text{thin}}$, $\sigma_{w,\text{thick}}$, $\sigma_{w,\text{gas}}$) and mid-plane matter densities ($\rho_{0,\text{thin}}$, $\rho_{0,\text{thick}}$, $\rho_{0,\text{gas}}$).

Given these specifications, the matter density of the respective components, here labelled by $x=\{\text{thin},\text{thick},\text{gas}\}$, follows the relation
\begin{equation}
    \rho(z)_x = \rho_{0,x} \exp \Bigg[ -\frac{\Phi(z)}{\sigma_{w,x}^2} \Bigg],
\end{equation}
where $\Phi(z)$ is found via the Poisson equation of Eq.~\eqref{eq:Poisson} and the total matter density, which is equal to
\begin{equation}
    \rho(z) = \rho(z)_\text{thin} + \rho(z)_\text{thick} + \rho(z)_\text{gas} + \rho_\text{DM}.
\end{equation}
By solving this system of equations, assuming boundary conditions $\Phi(0~\pc) = 0~\kmsecsq$ and $\partial \Phi/\partial z|_{z=0~\pc} = 0~\text{km}\,\text{s}^{-2}$, the phase-space distributions of the respective components are fully determined. 

The $10^5$ particles of our simulations were randomly drawn realisations of the stellar and gas steady state phase-space distributions, where the surface mass per particle is equal to $10^{-5}$ times the total surface mass of the gas and stellar components. When generating the data, meaning the two-dimensional histogram $\data_{i,j}$ representing observed stars, only the particles drawn from the thin and thick stellar disk components were used; particles belonging to the gas component were ignored because they would not be observed by the \emph{Gaia} survey.

\begin{figure}
        \includegraphics[width=1.\columnwidth]{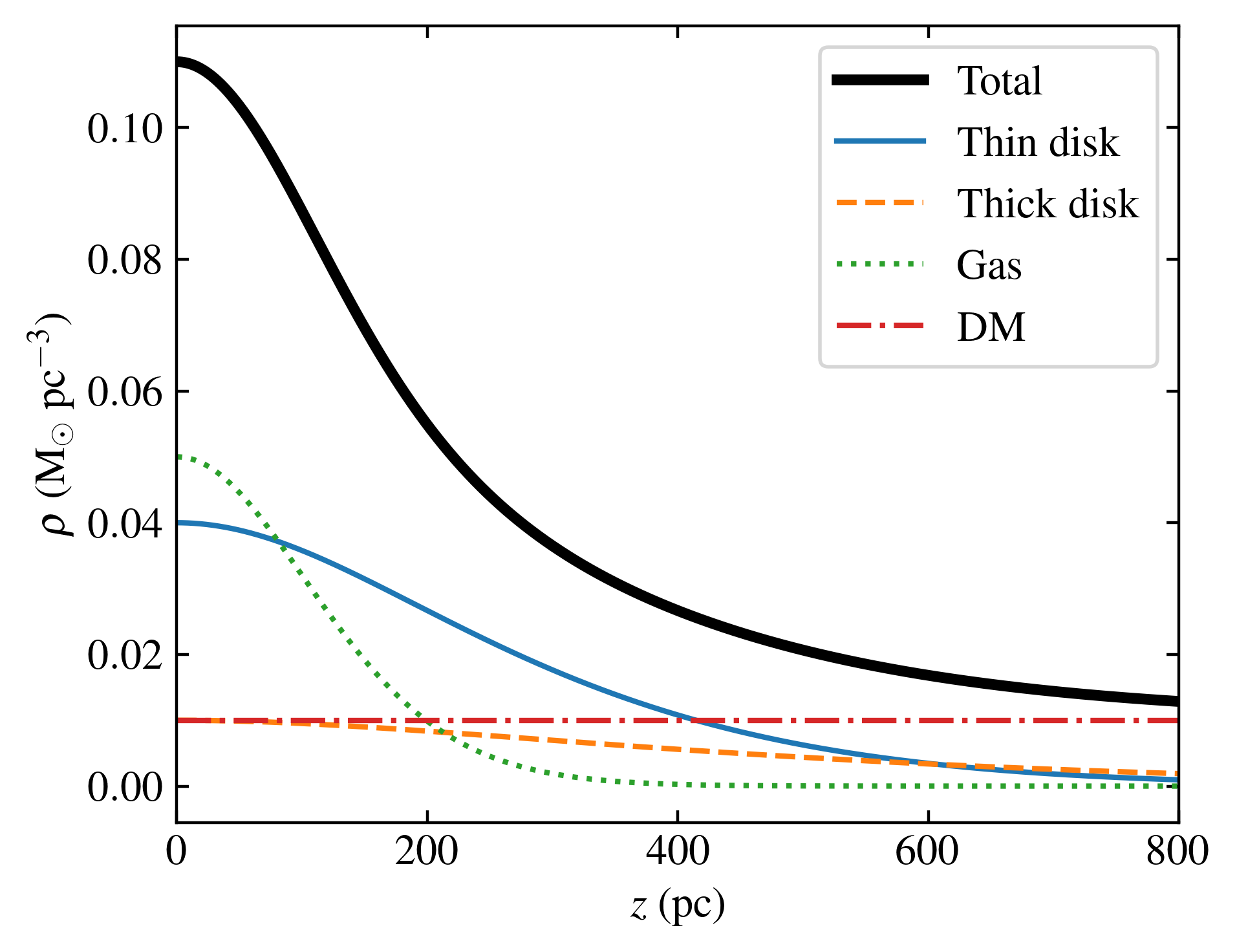}
    \caption{Example of the matter density distributions from which the particles of our one-dimensional simulations are drawn. In this plot, the distribution corresponds to that of simulation A; see Sect.~\ref{sec:simulations} and Table~\ref{tab:sim_parameters} for details.}
    \label{fig:initial_distributions}
\end{figure}

An example of the initial configuration of the matter density distributions is shown in Fig.~\ref{fig:initial_distributions}. The values for $\rho_\text{DM}$, $\sigma_{w,x}$, and $\rho_{0,x}$ correspond to those of simulation A, which are listed in Table~\ref{tab:sim_parameters}. The initial matter density distributions are mirror-symmetric with respect to the mid-plane ($z \rightarrow -z$).

\subsection{Perturbation}

After the initial set-up, we perturbed the simulation by applying an external force for a limited amount of time. The external force that created the perturbation is of the form
\begin{equation}\label{eq:satellite_perturbation}
    K_{z,\text{sat.}}(t,z) =
    -4 \pi G\, \Sigma \,
    \exp\Bigg( -\frac{t^2}{2\sigma_t^2} \Bigg) \,
    \tanh\Bigg[\frac{z-(z_{0,\text{sat.}}+w_\text{sat.}t)}{H}\Bigg],
\end{equation}
where $\Sigma$ is a surface mass, $\sigma_t$ corresponds to the duration of the external force, $H$ is a softening length, $z_{0,\text{sat.}}$ is the position of the satellite at time $t=0$, and $w_\text{sat.}$ is its constant vertical velocity. The perturbation is active for a range in time of $(-2.5\sigma_t,\,2.5\sigma_t)$. The time $t$ is defined with respect to the instance when this external force was maximal ($t=0~\yr$). The specific numerical values we used for the simulations in this work are provided in Sect.~\ref{sec:A_and_B_sims} and Table~\ref{tab:sim_parameters}.

This perturbation can be thought of as a massive satellite that passes through the Galactic disk at some angle of incidence. Due to its velocity parallel to the disk, it only affects a disk region for a limited amount of time. The softening length would correspond to the physical scale of the satellite.

We also added a scattering component, written as $s_n$, unique for each separate particle, randomly distributed according to a Gaussian centred on unity and with a standard deviation of 1/2. The particles are affected to different extents by the external force, according to
\begin{equation}\label{eq:scattering}
    K_{z,\text{sat.},n}(t,z) = s_n K_{z,\text{sat.}}(t,z).
\end{equation}
If such a scattering component is not included, the perturbation produces definite holes in the $(z,w)$-plane and creates a spiral structure that is too clearly defined. The scattering component was introduced in order to create a smoother distribution of over- and under-densities, similar to the actual phase-space spiral of the Milky Way.

The external force we used in this work was not meant to represent a fully realistic perturbation due to a passing satellite or other source; a fully realistic perturbation of the disk can only be simulated in a complete six-dimensional phase space. Rather, the aim of the simulations was to create phase-space spirals that are qualitatively similar to the actual phase-space spiral of the Milky Way. Furthermore, the relatively low vertical speed of the passing satellite\footnote{The satellite's vertical speed is low with respect to the expected total speed of a satellite at the Sun's position, where the rotational velocity is of the order of $230~\kmsec$. However, the satellite's vertical speed ($50$--$70~\kmsec$) is fairly high compared to the average speed of the stars (roughly $20~\kmsec$).}
was chosen in order to achieve a perturbation that affects stars at lower vertical energies and also produces a perturbation that is asymmetrical (see \citealt{2014MNRAS.440.1971W} for how a satellite's properties are correlated with the resulting disk perturbation). This gives rise to a spiral that is similar to that of the Milky Way in the following sense: While the spiral structure is not at all visible close to the origin of the $(z,w)$-plane, outside this innermost region the relative density of the spiral structure is close to constant in amplitude.

The main difficulty we had in reproducing the Milky Way phase-space spiral in the $(z,w)$-plane is related to its asymmetry. The actual phase-space spiral of the Milky Way has a single arm. Despite our efforts, we could only produce double-armed spirals, albeit with some asymmetry in terms of the relative amplitudes of the two arms. Even though the applied external force of our simulations was asymmetrical and produced clearly asymmetrical initial perturbations, the subsequent evolution of the perturbations evolved into double-armed spirals. This is further demonstrated in Appendix~\ref{app:double_arms}, where we run a simulation with a more idealised and clearly asymmetric initial perturbation; even in this case, a second arm develops spontaneously due to self-gravity after roughly $200~\Myr$. Perhaps a less idealised, noisier environment is enough to quench the formation of the secondary arm, possibly as a consequence of occupying a six-dimensional phase space or other effects of dynamical diffusion. Even though our simulations did not reproduce the single-armed spiral structure, we are still confident that this test on mock data is valuable as a proof of concept and demonstration of a completely novel method for inferring the Galactic gravitational potential.

\subsection{Simulations A and B}\label{sec:A_and_B_sims}

In this work we ran two separate simulations, referred to as simulation A and simulation B, on which we tested our analytical model. The parameters of these two simulations are listed in Table~\ref{tab:sim_parameters}.

In terms of the initial configuration of simulations A and B, their respective matter density distributions differ, both in terms of stars and gas. For the perturbation, simulations A and B differ in all parameters. The external force applied to simulation B was weaker (mainly due to a smaller surface mass $\Sigma$) and produced a spiral with a smaller relative amplitude. The perturbing satellites' total ranges in height are equal to $\sigma_t \times w_\text{sat.} \simeq \{2,3.6\}~\kpc$ for simulations A and B, respectively.

For simplicity, we set $Z_\odot=0~\pc$ and $W_\odot=0~\kmsec$ when producing the two-dimensional data histogram $\data_{i,j}$. In reality, the Sun's vertical position and velocity are non-zero, but this would be trivial to correct for as these values are known to sufficient precision (see for example \citealt{2020arXiv201202061G}) compared to the resolution at which we resolve the phase-space spiral in our method. In our method of inference, the parameters $Z_\odot$ and $W_\odot$ are still free parameters, which were fitted in the first step of our minimisation procedure.

{\renewcommand{\arraystretch}{1.6}
\begin{table}[ht]
        \centering
        \caption{Parameters of simulations A and B. The top seven parameters in this list determine the initial steady state configuration, and the bottom five parameters determine the properties of the perturbation.}
        \label{tab:sim_parameters}
    \begin{tabular}{| l | r | r |}
                \hline
                \backslashbox{Parameter}{Simulation} & A & B \\
                \hline
                $\rho_{0,\text{thin}}~(\Msunppcc)$ & $0.04$ & $0.02$ \\
                $\rho_{0,\text{thick}}~(\Msunppcc)$ & $0.01$ & $0.02$ \\
                $\rho_{0,\text{gas}}~(\Msunppcc)$ & $0.05$ & $0.04$ \\
                $\rho_\text{DM}~(\Msunppcc)$ & $0.01$ & $0.01$ \\
                $\sigma_{w,\text{thin}}~(\kmsec)$ & $16$ & $16$ \\
                $\sigma_{w,\text{thick}}~(\kmsec)$ & $24$ & $24$ \\
                $\sigma_{w,\text{gas}}~(\kmsec)$ & $8$ & $10$ \\
                $\Sigma~(\Msunppcsquare)$ & $6$ & $4$ \\
                $\sigma_t~(\Myr)$ & $40$ & $50$ \\
                $H~(\pc)$ & $50$ & $80$ \\
                $z_{0,\text{sat.}}~(\pc)$ & $600$ & $-400$ \\
                $w_\text{sat.}~(\kmsec)$ & $-50$ & $-70$ \\
        \hline
        \end{tabular}
\end{table}}

\section{Results}\label{sec:results}

In this section we present the results of simulations A and B. Our method was applied to each simulation at two separate times. For simulation A, these times are $t_{A,1}=400~\Myr$ and $t_{A,2}=600~\Myr$; for simulation B, they are $t_{B,1}=400~\Myr$ and $t_{B,2}=500~\Myr$. For these instances in time, the spiral winding is roughly one complete period in the spiral region of our analytical model, where the outer boundary is defined by the mask function $M(Z,W)$ and the inner boundary is defined by the model's lower energy bound $m(z,w\,|\,\popp_\text{spiral})$ (see Eqs.~\eqref{eq:mask} and \eqref{eq:inner_boundary}). For the bulk density distribution $B(z,w,\, | \, \popp_\text{bulk})$, which is minimised in the first step of the fitting procedure, we used a Gaussian mixture model consisting of $K=6$ Gaussians; fitting a higher number of Gaussians did not significantly improve the maximum likelihood.

\begin{figure*}
        \includegraphics[width=1.\textwidth]{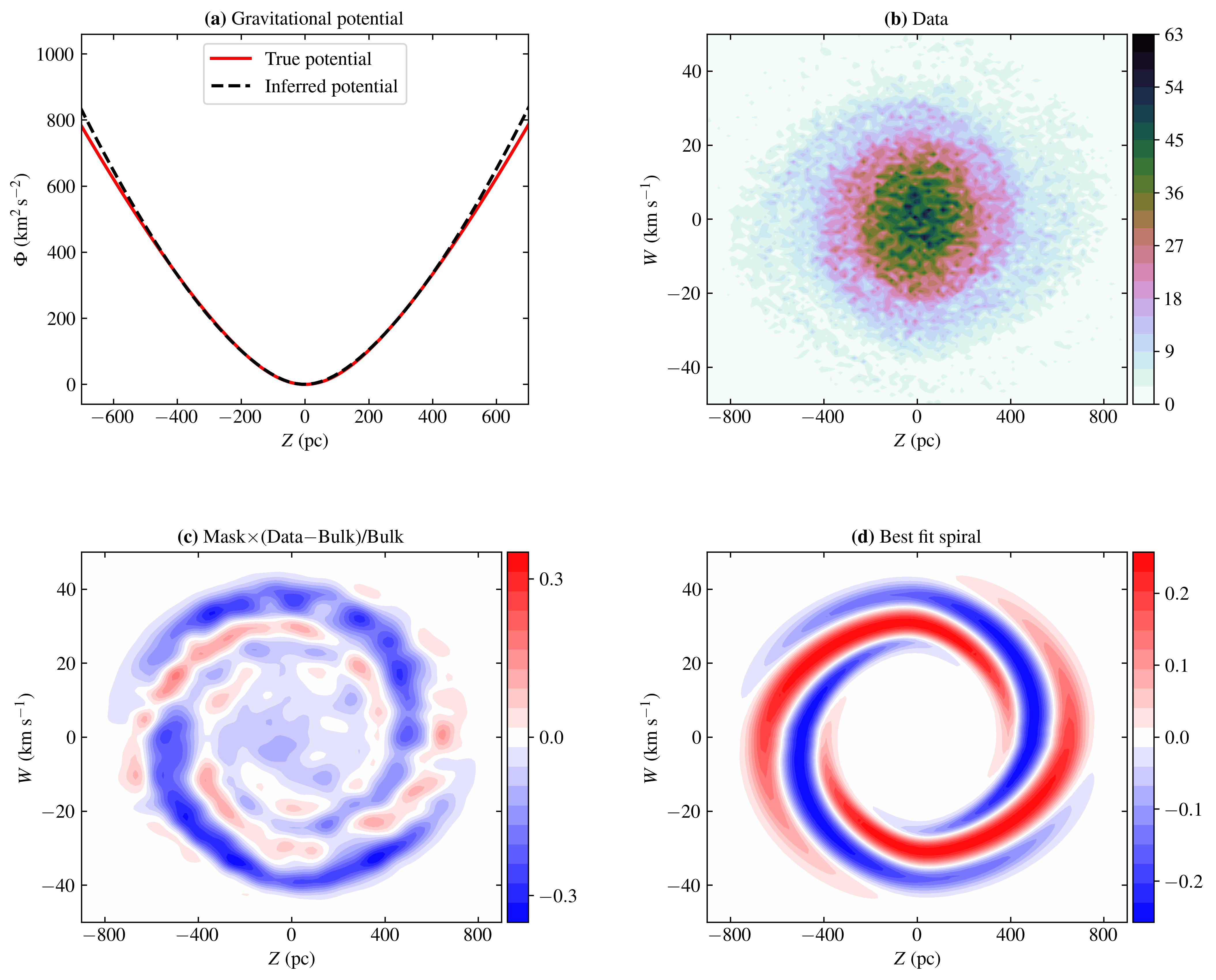}
    \caption{Data and results of simulation A at time $t_{A,1}=400~\Myr$. The four panels show: \textbf{(a)} the true and inferred gravitational potentials; \textbf{(b)} the two-dimensional data histogram $\data_{i,j}$ in the $(Z,W)$-plane; \textbf{(c)} the phase-space spiral as extracted from the data (see Eq.~\eqref{eq:pandel_c} for the precise definition) and smoothed to an effective bin size of $(40~\pc)\times(2~\kmsec)$;  and \textbf{(d)} the best fit spiral density $S(Z+Z_\odot,W+W_\odot \, | \, \popp_\text{spiral})$, as defined in Eq.~\eqref{eq:pandel_d}. In panel \textbf{(a)}, the height is plotted in the range $Z \in [-700,700]~\pc$, corresponding to the outer limit of the mask function; in all other panels, the range is $Z \in [-900,900]~\pc$. In panels \textbf{(c)} and \textbf{(d)}, the outer boundary corresponds to that of the mask function of Eq.~\eqref{eq:mask}; in panel \textbf{(d)}, the inner boundary corresponds to the lower limit in vertical energy for the spiral model, according to Eq.~\eqref{eq:inner_boundary}.}
    \label{fig:simA_400Myr}
\end{figure*}

\begin{figure*}
        \includegraphics[width=1.\textwidth]{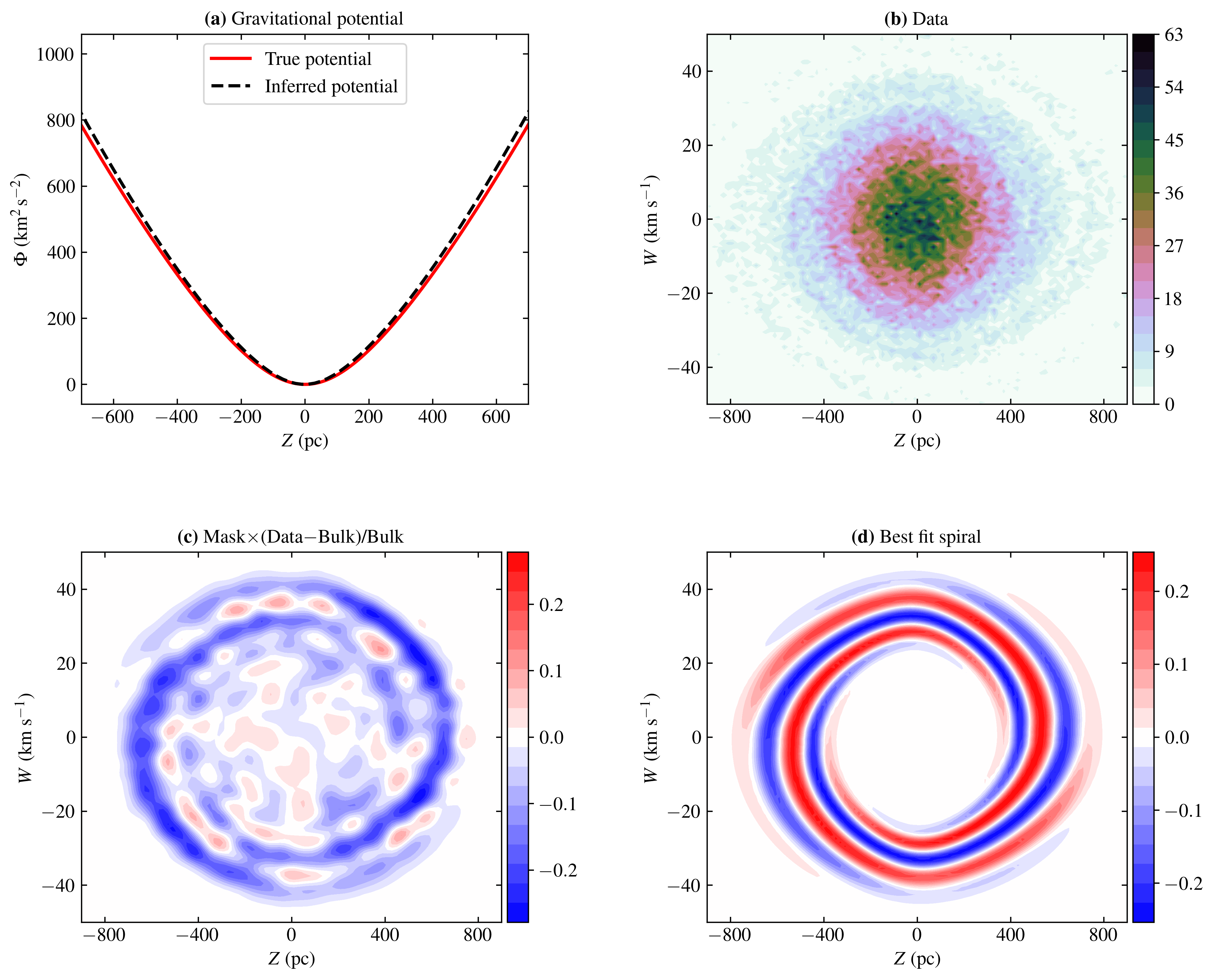}
    \caption{Same as Fig. \ref{fig:simA_400Myr}, but for simulation A at time $t_{A,2}=600~\Myr$.}
    \label{fig:simA_600Myr}
\end{figure*}

\begin{figure*}
        \includegraphics[width=1.\textwidth]{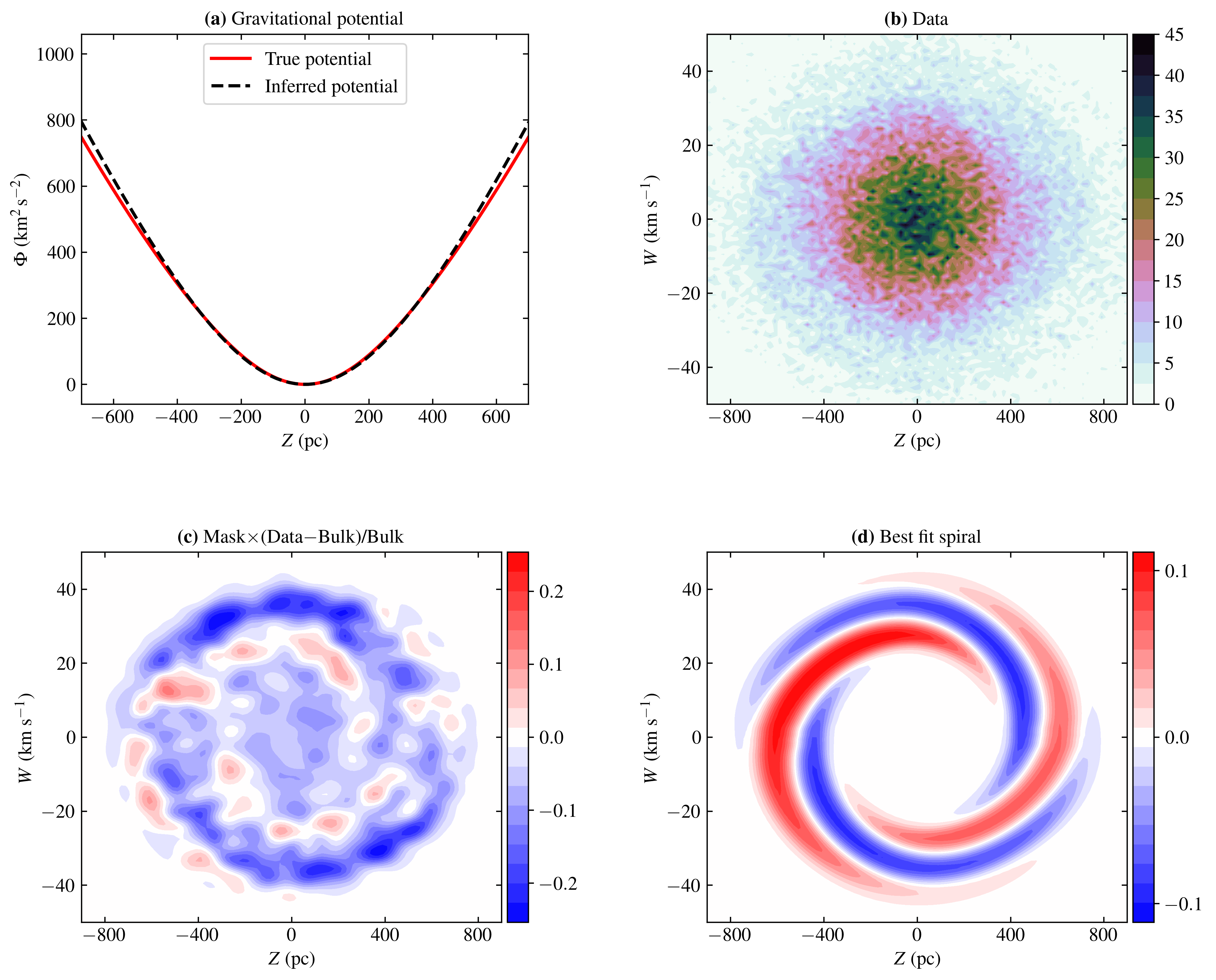}
    \caption{Same as Fig. \ref{fig:simA_400Myr}, but for simulation B at time $t_{B,1}=400~\Myr$.}
    \label{fig:simB_400Myr}
\end{figure*}

\begin{figure*}
        \includegraphics[width=1.\textwidth]{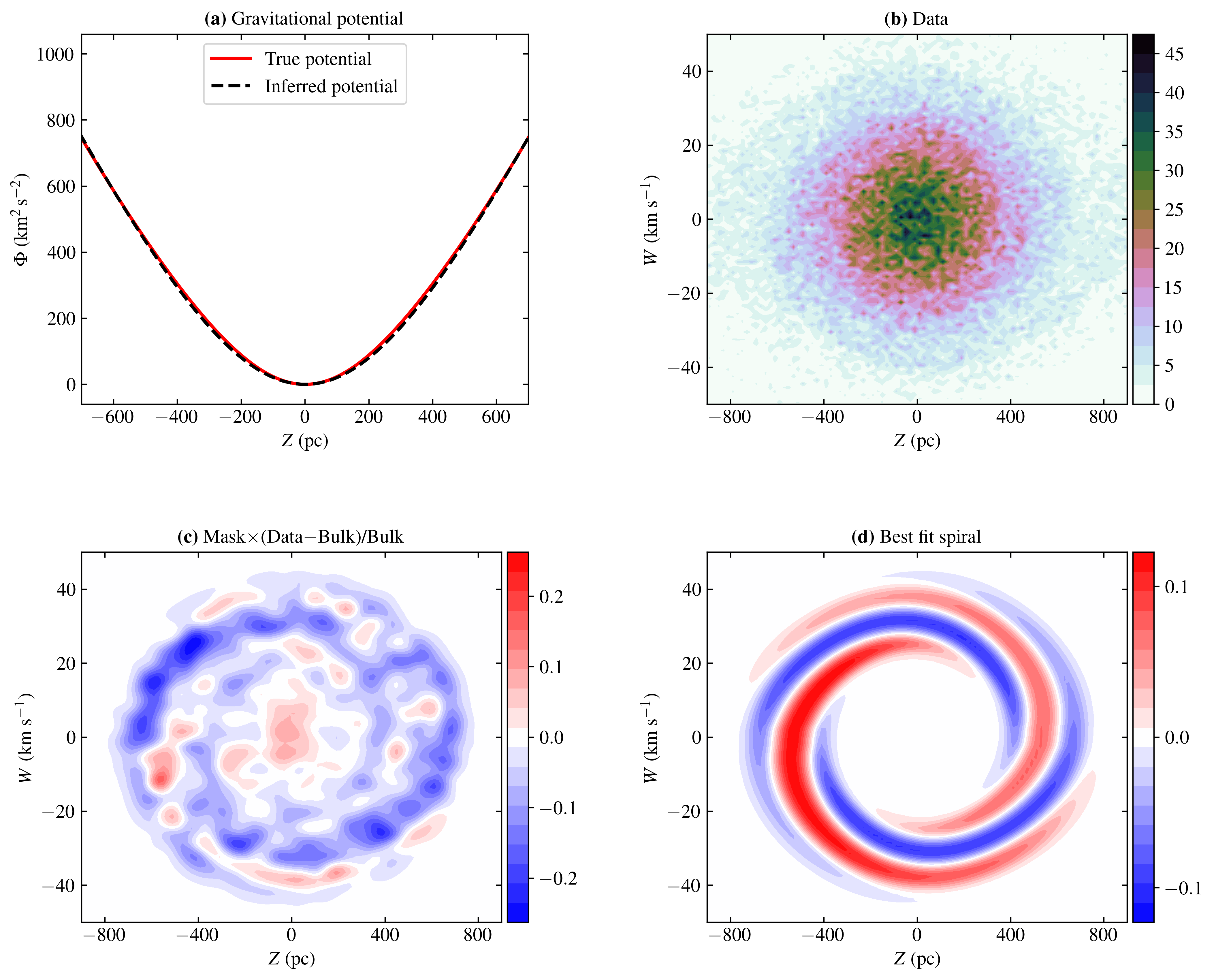}
    \caption{Same as Fig. \ref{fig:simA_400Myr}, but for simulation B at time $t_{B,2}=500~\Myr$.}
    \label{fig:simB_500Myr}
\end{figure*}

The results of the four different cases are presented in Figs.~\ref{fig:simA_400Myr}--\ref{fig:simB_500Myr}. The figures each contain four panels that show the true and inferred gravitational potential, a two-dimensional histogram of the data, the spiral extracted directly from the data, and the best fit spiral of our analytical model. The spiral as extracted from the data, seen in panel \textbf{(c),} is defined as
\begin{equation}\label{eq:pandel_c}
    M(Z_i,W_j)\times\bigg[\frac{\data_{i,j}-B(Z_i+Z_\odot,W_j+W_\odot \, | \, \popp_\text{bulk})}{B(Z_i+Z_\odot,W_j+W_\odot \, | \, \popp_\text{bulk})}\bigg].
\end{equation}
In panel \textbf{(d)}, the spiral of our analytical model also includes the lower energy bound of Eq.~\eqref{eq:inner_boundary}, according to
\begin{equation}\label{eq:pandel_d}
\begin{split}
    & M(Z_i,W_j) \times m(Z_i+Z_\odot,W_j+W_\odot \, | \, \rho_h) \\
    & \times S(Z_i+Z_\odot,W_j+W_\odot \, | \, \popp_\text{spiral}).
\end{split}
\end{equation}

We only show our inferred results in terms of the best fit, meaning the maximised likelihood. Running the minimisation algorithm is already very computationally intensive (see Sect.~\ref{sec:implementation} for details), and computing a full Markov chain Monte Carlo would be very costly and without much return on investment because our results are dominated by systematic rather than statistical uncertainties; because of the high number of stars and the simplicity of our analytical model, the statistical uncertainty is very small (below the 1 percent level). When applying our model to the real phase-space spiral of the Milky Way, the total amount of available stars will be even higher and there will be plenty of other potential sources of systematic bias; in other words, applying this method to real data will most probably only make systematic effects dominate even more.

The inferred potential at greater heights ($|z| \gtrsim 600~\pc$) is less robustly inferred and in fact largely an extrapolation with regards to its shape at smaller heights. The main reason for this is that only a very small portion of the two-dimensional histogram, after applying the mask of Eq.~\eqref{eq:mask}, covers such heights. There seems to be a slight bias towards a steeper gravitational potential in the regime of larger $|z|$. In terms of accuracy, the most robustly inferred quantity is the gravitational potential value at heights around 400 to 500 pc, for which the relative error is only a few percent (strictly less than $5~\%$). The precise shape of the gravitational potential close to the Galactic mid-plane is less robustly inferred; as was demonstrated in Sect.~\ref{sec:examples} and Fig.~\ref{fig:schematic}, the shape of the gravitational potential at low heights gives rise to quite small differences in the spiral's shape.

In terms of the time that has passed since the perturbation (the parameter $t$), the best fit results are $t_{A,1}=432~\Myr$, $t_{A,2}=506~\Myr$, $t_{B,1}=334~\Myr$, and $t_{B,2}=405~\Myr$. The true values are $400~\Myr$, $600~\Myr$, $400~\Myr$, and $500~\Myr$, respectively, where the true time is defined in terms of when the external force of Eq.~\eqref{eq:satellite_perturbation} was maximal. However, the force is actually active for roughly $100~\Myr$ both before and after $t=0~\Myr$. Hence, we did not expect the inferred value for $t$ to be accurate within such time frames.

In order to demonstrate the limits of our model of inference, we also show results for simulation B at times $t_{B,3}=600~\Myr$ and $t_{B,4}=700~\Myr$ in Appendix~\ref{app:results}. At these times, the spiral arms can no longer be seen by eye as clear continuous structures and our method starts to lose accuracy in terms of the inferred gravitational potential (the relative error is as high as 10~\% for the inferred potential at $|z|=400~\pc$). We also ran other similar simulations, for example with more complicated initial stellar phase-space distributions, for which we obtained similar results.

\section{Discussion}\label{sec:discussion}

When applying our method of inference to one-dimensional simulations, we were able to infer the gravitational potential to high accuracy. Our method extracts information from the time-varying structure of a phase-space spiral; this is a novel approach, constituting a break from traditional methods that are based on the assumption of a steady state. For this reason, our method is also complementary to such traditional methods and likely subject to different sources of systematic bias. We note that there are also other methods for extracting information from the phase-space spiral. For example, \cite{2021arXiv210107080L} infer the gravitational potential of the disk using the bulk stellar density in the $(z,w)$-plane and then fit the spiral to the residual of the inferred steady state distribution. The fitted spiral constitutes a consistency check of the inferred potential and is seen as a straight line in the plane of angle and frequency.

While our method is accurate in terms of one-dimensional simulations, the actual Milky Way is significantly more complex. To begin with, stars in the Galactic disk do not only oscillate in the vertical direction, but also move in the radial direction as they orbit the Galactic centre. Indeed, the phase-space spiral of the Milky Way has a different form depending on cuts in angular momentum or the galactocentric radius (see for example Fig. 14 in \citealt{2019MNRAS.485.3134L}, as well as \citealt{2019MNRAS.489.4962K}). When applying our method to \emph{Gaia} data for the first time, we plan to select stars of the solar neighbourhood that have approximately circular orbits, for which the assumptions of our model are most likely to be valid. It would be interesting, although probably quite complicated, to see what kind of information about the global structure of the Milky Way could be extracted from the shape of the spiral in different bins of Galactic angular momentum. Ultimately, we would want to test and refine our method using a high-resolution simulation of a galaxy in three spatial dimensions. Up until now, simulations of Milky Way-like systems (such as \citealt{2018MNRAS.481..286L} and \citealt{2020arXiv200902434B}) have not had sufficient resolution to produce well-resolved phase-space spirals in stellar density in the $(z,w)$-plane. To do so requires billion-particle simulations, which have only recently become feasible (\citealt{2020MNRAS.499.2416A}, Hunt et al. in preparation). Furthermore, our modelling of the gravitational potential needs to be expanded somewhat; the Poisson equation is no longer one-dimensional but also has a smaller contribution from the radial direction (the so-called rotation curve term).

A useful test in terms of systematic uncertainties would be to perform separate analyses of different segments of the phase-space spiral. The assumptions of our analytical model are probably only valid over a certain range in vertical energy; fitting the spiral over a large area in the ${z,w}$-plane would give a small statistical uncertainty, but potentially at the price of large systematic errors. It would be possible, at least to some extent, to quantify systematic errors by comparing the results of the smaller, separate spiral segments. Another test in a similar vein would be to abandon the form of the function $\tilde{\varphi}(t,E_z)$ as defined in Eq.~\eqref{eq:angle_of_time}, which might not be completely accurate due to effects of, for example, self-gravity; instead, we could simply assume that the spiral angle $\tilde{\varphi}$ is a smooth function with respect to $E_z$ and model it as a Gaussian process.

Observational uncertainties are probably not a significant factor when applying this method to \emph{Gaia} data and our own Galaxy. The most significant observational uncertainties are associated with the parallax measurements. With \emph{Gaia} EDR3, those uncertainties are of the order of 0.02--0.03 mas for stars with absolute $G$-band magnitudes smaller than 15 \citep{2020arXiv201203380L}. Within kiloparsec distances, this corresponds to a relative uncertainty of at most a few percent (meaning at most a few tens of parsecs in distance). The uncertainties with respect to velocities are typically only a few hundred metres per second \citep{rv_systematics}. Neither distance- nor velocity-related uncertainties should be significant issues given the scale of the phase-space spiral and the large amount of available statistics.

Observing the phase-space spiral using \emph{Gaia} requires radial velocity measurements, which are not available for all stars. Cleaning the data to exclude stars with missing radial velocities introduces a significant selection effect. However, selection effects should not be a hindrance to our method, at least not as long as the selection function is fairly smooth. What is crucial is that the shape of the phase-space spiral be extracted with accuracy; any selection effects that vary smoothly with distance will be accounted for when subtracting the bulk density component.

\section{Conclusion}\label{sec:conclusion}

In this work we have developed a method for inferring the gravitational potential of the Galactic disk from the shape of a phase-space spiral in the $(z,w)$-plane. Our analytical model of the spiral, which is fitted to data, is based on a few simplifying assumptions, such as neglecting the self-gravity of the spiral and assuming that it evolves in a static potential. We tested our method on one-dimensional simulations (which do not adhere to the model assumptions) and were able to recover the gravitational potential with high accuracy. In cases where the phase-space spiral could be seen as a smooth continuous structure in the modelled region of the $(z,w)$-plane (similar to the actual spiral of the Milky Way), the relative error for the inferred potential at heights $|z| \simeq 500~\pc$ was as small as a few percent; poor accuracy, meaning relative errors of $\text{6--10}~\%$, was obtained only in cases where the phase-space spiral lacked these characteristics.

This article is the first in a series in which we plan to apply our method to the real Milky Way phase-space spiral, as well as test and refine our method using more complex Galaxy simulations. This is a step in the direction of modelling the time-dependent dynamics of the Milky Way and demonstrates that a time-varying structure is not necessarily an obstacle to dynamical mass measurements, but can in fact be an asset from which it is possible to extract useful information. The method in this paper completely disregards the bulk density distribution and is therefore complementary to traditional methods for which the bulk is the key quantity. Eventually, we would like to combine these separate methods (as also discussed by \citealt{2021arXiv210107080L}) and infer the gravitational potential of the Galactic disk using the joint information of the spiral and the bulk phase-space density distribution.

\begin{acknowledgements}
We would like to thank the anonymous referee for a useful and instructive review.
AW acknowledges support from the Carlsberg Foundation via a Semper Ardens grant (CF15-0384).
CL acknowledges funding from the European Research Council (ERC) under the European Union's Horizon 2020 research and innovation programme (grant agreement No. 852839).
PFdS acknowledges support by the Vetenskapsr{\aa}det (Swedish Research Council) through contract No. 638-2013-8993 and the Oskar Klein Centre for Cosmoparticle Physics.
This work made use of an HPC facility funded by a grant from VILLUM FONDEN (projectnumber 16599).
This work was supported in part by World Premier International Research Center Initiative (WPI Initiative), MEXT, Japan.

This research utilised the following open-source Python packages: \textsc{Matplotlib} \citep{matplotlib}, \textsc{NumPy} \citep{numpy}, \textsc{SciPy} \citep{scipy}, \textsc{Pandas} \citep{pandas}, \textsc{TensorFlow} \citep{tensorflow2015-whitepaper}.
\end{acknowledgements}




\bibliographystyle{aa} 
\bibliography{thisbib} 

\begin{appendix}

\section{Second arm forms via self-gravity}\label{app:double_arms}

In this section we demonstrate that a phase-space spiral develops double arms via self-gravity, even when only a single arm is present to begin with. We ran a simulation whose initial phase-space configuration was the same as for simulation A (see Sect.~\ref{sec:A_and_B_sims}) and then perturbed it by hand at time $t=0$ by adding a kick to the vertical velocity, written as $w^+$, for a subset of the simulation particles. This was done according to the formula
\begin{equation}
    w^+ = \Theta(z) \times \sigm\Bigg[\frac{w-15~\kmsec}{10~\kmsec}\Bigg] \times 10~\kmsec,
\end{equation}
where $\Theta$ is the Heaviside step function. In this manner, only simulation particles with positive $z$ and positive $w$ were affected by the perturbation.

\begin{figure}
        \includegraphics[width=1.\columnwidth]{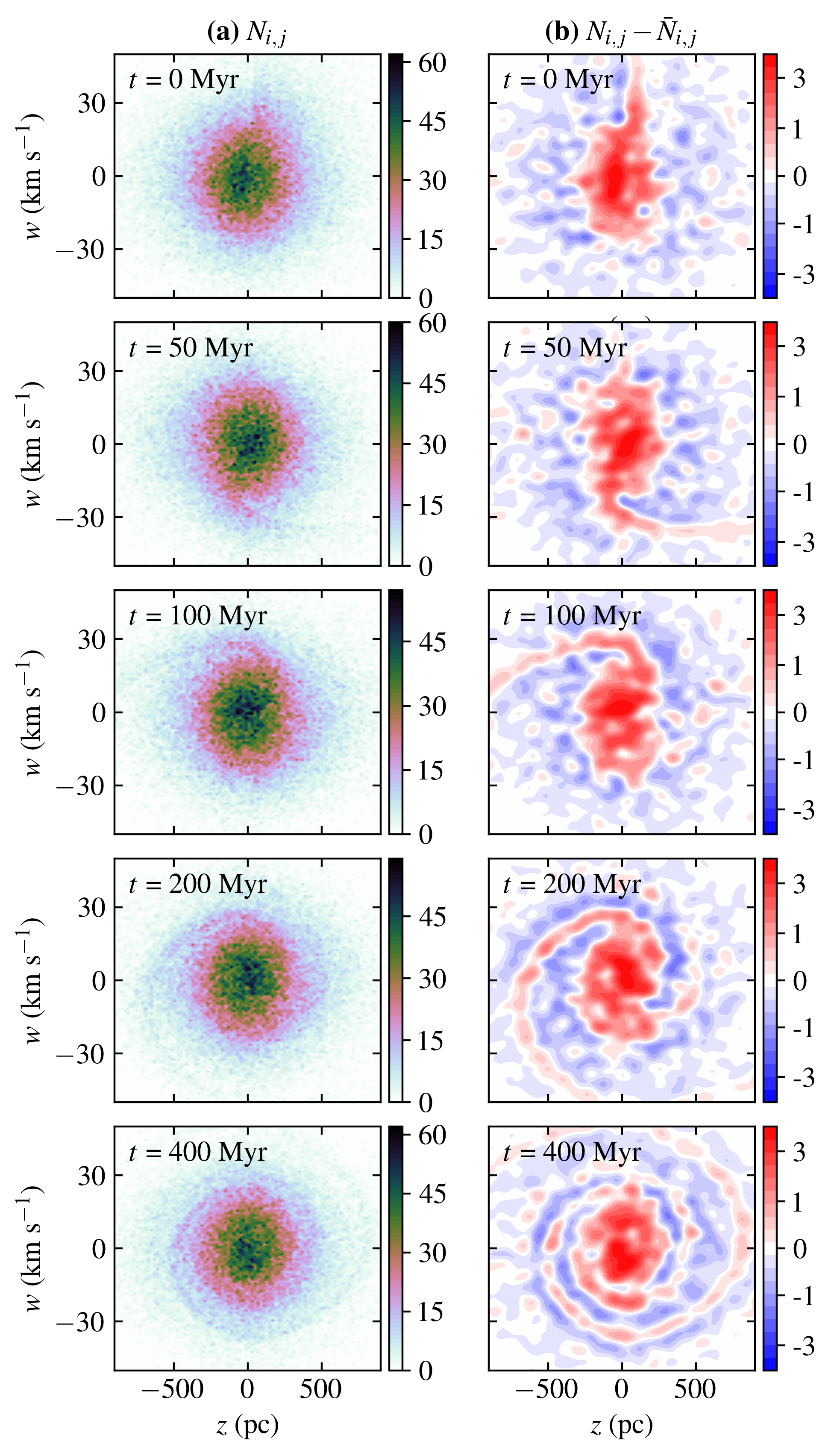}
    \caption{Stellar number density distribution in the $(z,w)$-plane shown at times $t=\{0,50,100,200,400\}$ Myr (going from top to bottom) for a simulation affected by an asymmetric perturbation at time $t=0~\Myr$. The left and right panels show \textbf{(a)} a histogram of the total stellar density with bin size $(20~\pc)\times(1~\kmsec)$; and \textbf{(b)} the difference between the stellar density histogram and the average of the surrounding histogram bins. All panels have the same coordinate axis. The colour scheme in the panels on the right-hand side is not linear, but in $\text{arcsinh}$ scale.}
    \label{fig:double_arm_forms}
\end{figure}

After this initial perturbation, the phase-space distribution evolves under self-gravity according to the vertical acceleration of Eq.~\eqref{eq:vertical_force}. The distribution of stars in the $(z,w)$-plane is shown at different times in Fig.~\ref{fig:double_arm_forms}. The histograms on the left-hand side, labelled $N_{i,j}$, show the stellar number density in bins with size $(20~\pc)\times(1~\kmsec)$. The panels on the right-hand side show the difference between the stellar number density histogram and the average value of its neighbouring area; this is written as $N_{i,j}-\bar{N}_{i,j}$, where $\bar{N}_{i,j}$ is equal to $N_{i,j}$ convoluted with a bivariate Gaussian distribution with standard deviations of $80~\pc$ and $4~\kmsec$. For better visibility, the panels on the right-hand side are illustrated with a resolution of $(40~\pc)\times(2~\kmsec)$.

Even though the initial perturbation is asymmetric and only a single spiral arm is visible at $t=100~\Myr$, a second arm develops; it is visible at $t=200~\Myr$ and becomes even more visible  at $t=400~\Myr$. This indicates that a symmetric spiral component (i.e. a double-armed spiral) develops spontaneously as an effect of self-gravity in our one-dimensional simulations. We tried many other initial simulation conditions, for example: varying the matter components' respective scale heights and mid-plane densities; varying the satellite perturbation's properties (most importantly, the vertical speed) in order to perturb stars with either higher or lower vertical energies and to produce more of a bending or breathing mode; and trying a variety of more idealised and strictly asymmetrical perturbations (similar to the one shown in Fig.~\ref{fig:double_arm_forms}). Despite these efforts, we always found similar results, which tended towards a symmetric phase-space spiral.

\section{Inferred vertical acceleration and total matter density}\label{sec:inferred_Kz_and_rho}

\begin{figure*}
        \includegraphics[width=1.\textwidth]{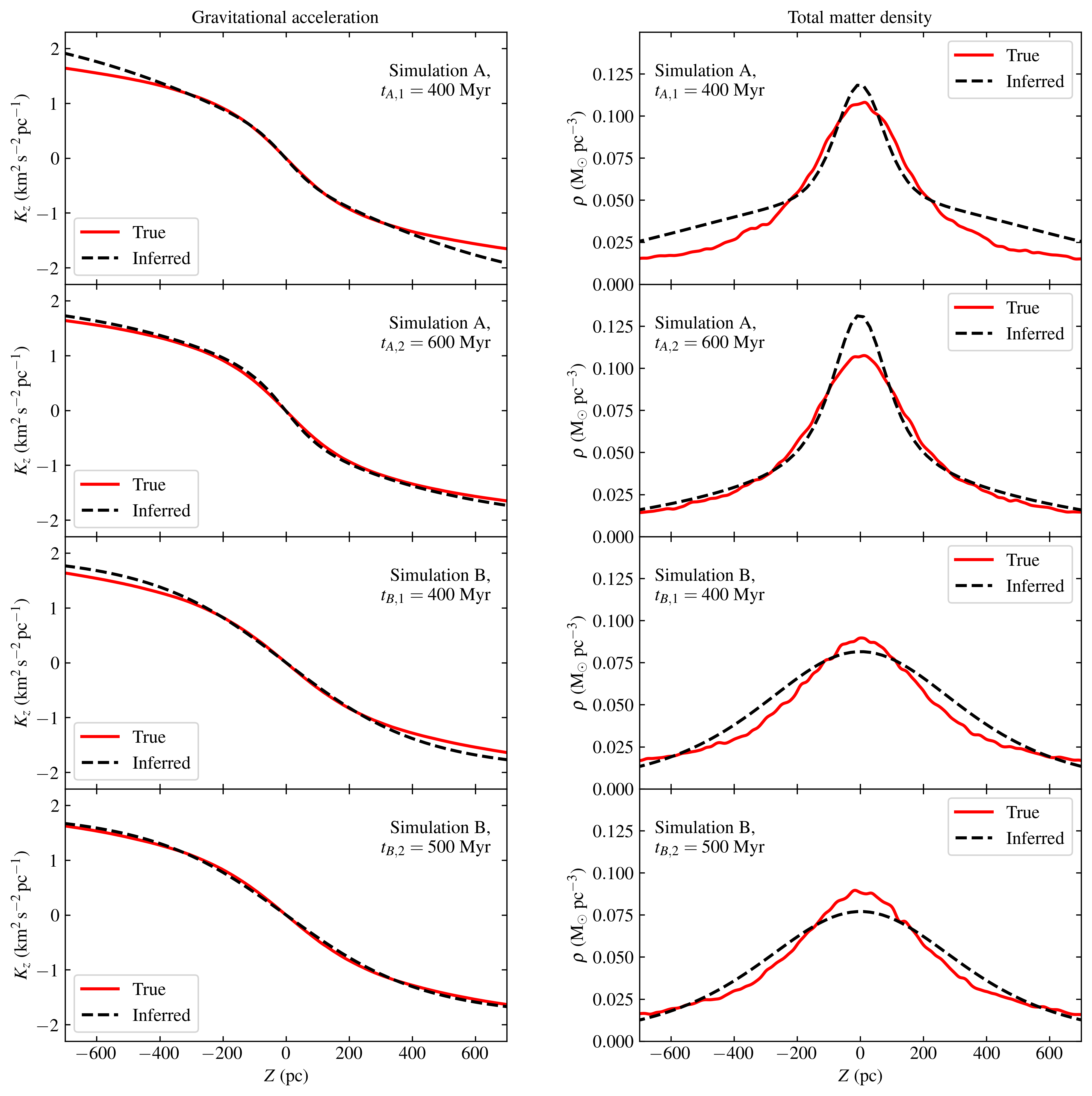}
    \caption{Inferred vertical acceleration, $K_z(Z)$, and total matter density, $\rho(Z)$, for simulations A and B. All panels share the same horizontal axis.}
    \label{fig:inferred_Kz_and_rho}
\end{figure*}

In Fig.~\ref{fig:inferred_Kz_and_rho} we show the inferred vertical acceleration ($K_z = -\partial \Phi / \partial Z$) and the inferred total matter density ($\rho$, fulfilling the Poisson equation; see Eq.~\eqref{eq:Poisson}). These quantities are proportional to the first and second derivatives of the gravitational potential. For this reason, they are less robustly inferred, and the matter density is especially sensitive to the precise shape of the inferred gravitational potential.

In the inferred matter density profiles, the scale height, shape, and mid-plane matter density are not very accurately inferred quantities.
However, the gravitational acceleration for heights $|Z|$ in the range of 250--300 pc (or, equivalently, the total surface matter density integrated over heights $|Z|<300~\pc$) is accurately inferred for all four examples, with a relative error strictly smaller than 4~\%. The gravitational acceleration and matter density at greater heights ($|Z|\gtrsim 400~\pc$) are not very accurately inferred, which is in large part due to the small amount of statistics at these heights (connected to how the gravitational potential is poorly inferred for $|Z|\gtrsim 600~\pc$; see Sect.~\ref{sec:results} for further details).

\section{Further results for simulation B}\label{app:results}

\begin{figure*}
        \includegraphics[width=1.\textwidth]{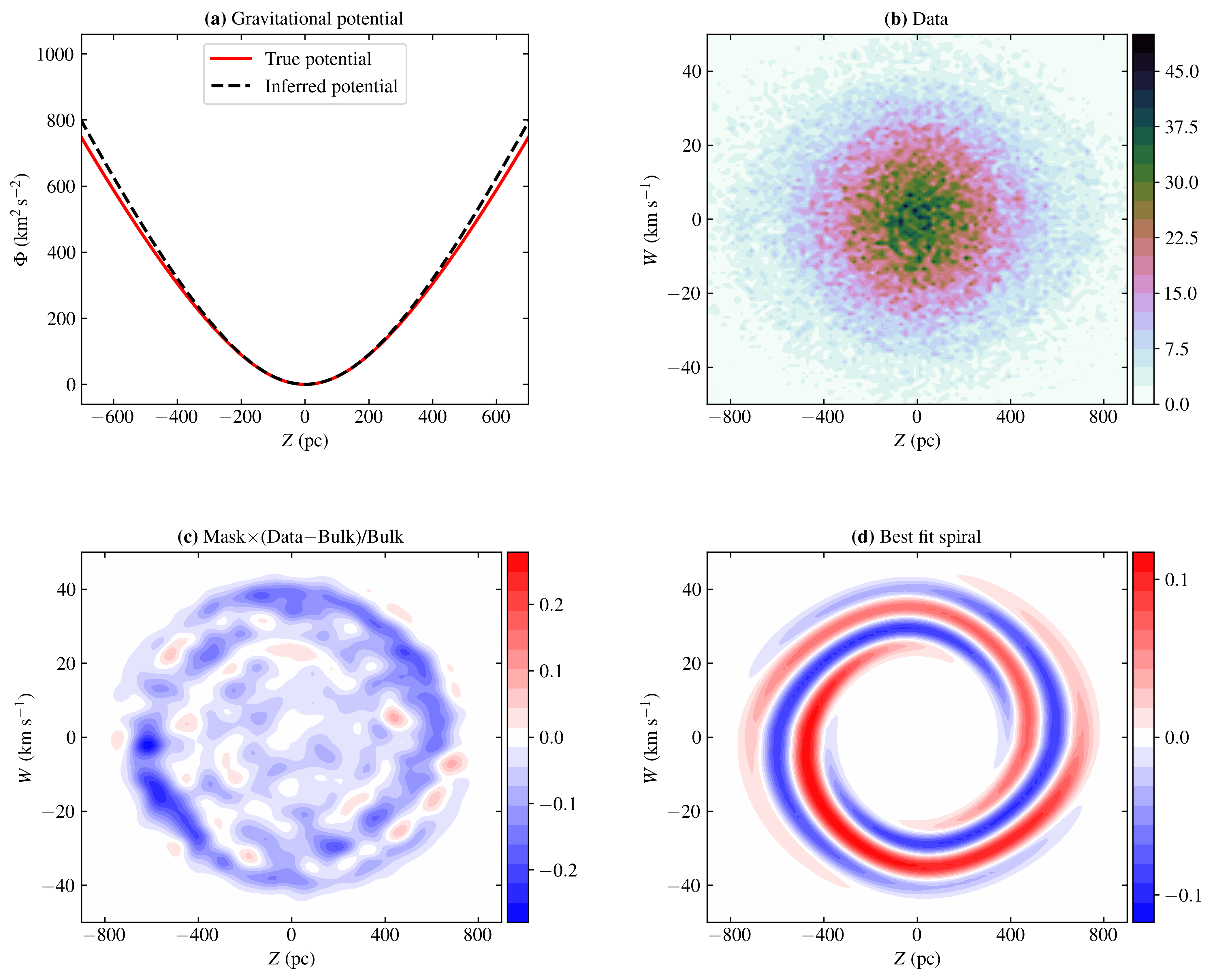}
    \caption{Same as Fig. \ref{fig:simA_400Myr}, but for simulation B at time $t_{B,3}=600~\Myr$.}
    \label{fig:simB_600Myr}
\end{figure*}

\begin{figure*}
        \includegraphics[width=1.\textwidth]{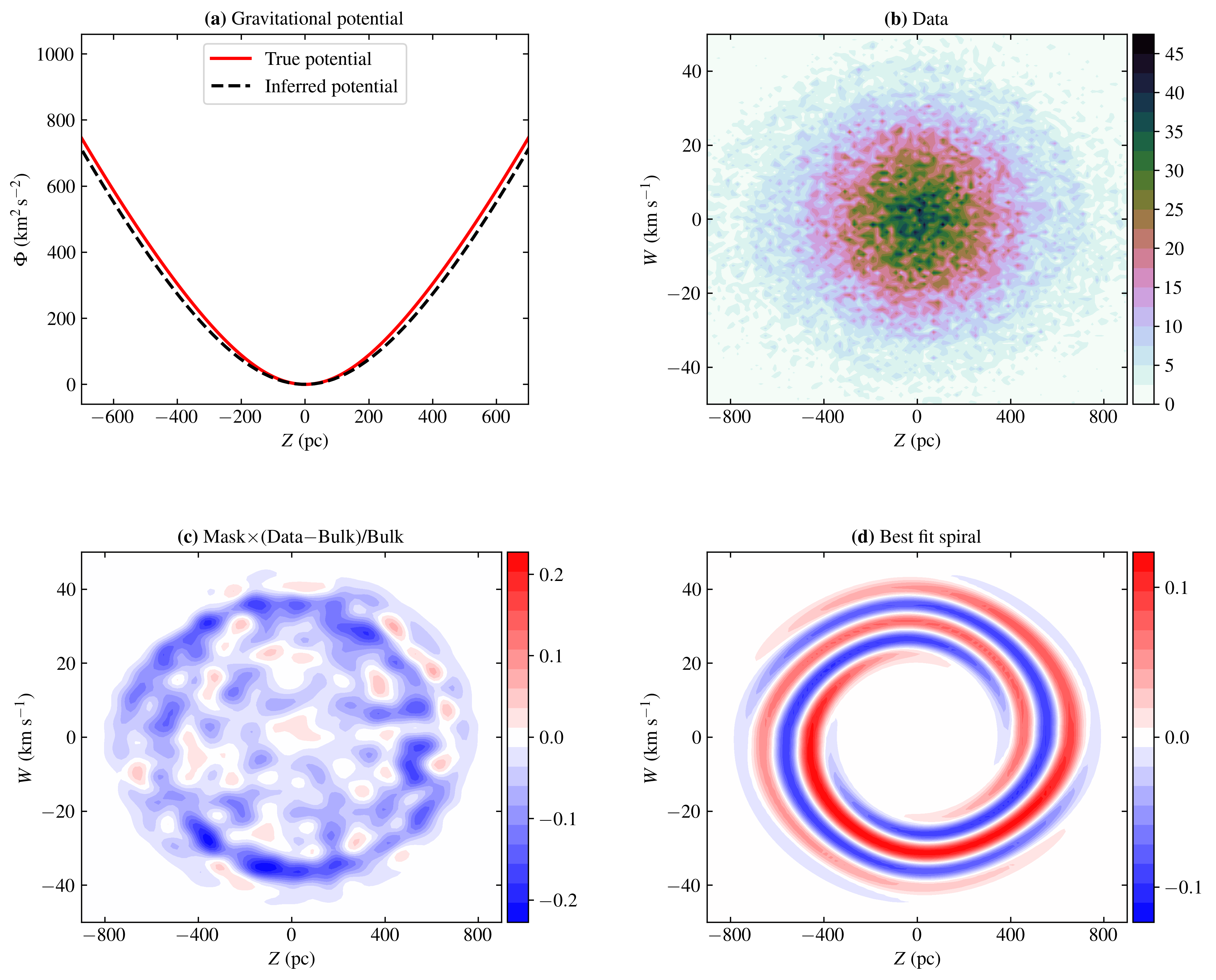}
    \caption{Same as Fig. \ref{fig:simA_400Myr}, but for simulation B at time $t_{B,4}=700~\Myr$.}
    \label{fig:simB_700Myr}
\end{figure*}

In this section we present some further results for simulation B, at times $t_{B,3}=600~\Myr$ and $t_{B,4}=700~\Myr$. We do so in order to demonstrate the limits of our method. Simulation B was subjected to a relatively weak perturbation, and the resulting spiral has a low amplitude (roughly $10~\%$ in terms of relative stellar number density, compared to roughly $20~\%$ for simulation A). With time and further winding, the spiral structure becomes all the more phase-mixed and increasingly hard to detect, eventually disappearing completely. In Figs.~\ref{fig:simB_600Myr} and \ref{fig:simB_700Myr}, we show results in the interim, where the spiral begins to be more difficult to see.

At these later times, when the spiral structure is less pronounced, our method loses accuracy. In terms of the gravitational potential at heights in the range 400 to 500 pc, the relative error is roughly 6~\% for $t_{B,3}$ and 10~\% for $t_{B,4}$. The inferred times since the perturbation are $t_{B,3}=451~\Myr$ and $t_{B,4}=725~\Myr$ (true values are $600~\Myr$ and $700~\Myr$).

\end{appendix}

\end{document}